\documentclass[10pt]{book}
\usepackage{array}
\usepackage{graphicx}
\usepackage{multirow}
\usepackage[T1]{fontenc}
\usepackage{amsthm,amsmath,amssymb}

\begin{document}


\renewcommand{\baselinestretch}{2}


\markboth{\hfill{\footnotesize\rm Øystein Sørensen, Arnoldo Frigessi and Magne Thoresen} \hfill}
{\hfill {\footnotesize\rm Measurement Error in Lasso} \hfill}

\renewcommand{\thefootnote}{}
$\ $\par


\fontsize{10.95}{14pt plus.8pt minus .6pt}\selectfont
\vspace{0.8pc}
\centerline{\large\bf MEASUREMENT ERROR IN LASSO:}
\vspace{2pt}
\centerline{\large\bf IMPACT AND LIKELIHOOD BIAS CORRECTION}
\vspace{.4cm}
\centerline{Øystein Sørensen, Arnoldo Frigessi and Magne Thoresen}
\vspace{.4cm}
\centerline{\it University of Oslo}
\vspace{.55cm}
\fontsize{9}{11.5pt plus.8pt minus .6pt}\selectfont


\begin{quotation}
\noindent {\it Abstract:}
Regression with the lasso penalty is a popular tool for performing dimension reduction when the number of covariates is large. In many applications of the lasso, like in genomics, covariates are subject to measurement error. We study the impact of measurement error on linear regression with the lasso penalty, both analytically and in simulation experiments. A simple method of correction for measurement error in the lasso is then considered. In the large sample limit, the corrected lasso yields sign consistent covariate selection under conditions very similar to the lasso with perfect measurements, whereas the uncorrected lasso requires much more stringent conditions on the covariance structure of the data. Finally, we suggest methods to correct for measurement error in generalized linear models with the lasso penalty, which we study empirically in simulation experiments with logistic regression, and also apply to a classification problem with microarray data. We see that the corrected lasso selects less false positives than the standard lasso, at a similar level of true positives. The corrected lasso can therefore be used to obtain more conservative covariate selection in genomic analysis.\par

\vspace{9pt}
\par
\end{quotation}\par


\fontsize{10.95}{14pt plus.8pt minus .6pt}\selectfont

\setcounter{chapter}{1}
\setcounter{equation}{0} 
\noindent {\bf 1. Introduction}

Due to rapid technological progress, complex, high-dimensional data sets are now commonplace in a range of fields, e.g., genomics and finance. Various penalization schemes have been proposed, which shrink the parameter space, including the Dantzig selector (Candès and Tao (2007)), the lasso (Tibshirani (1996)), ridge regression (Hoerl and Kennard (1970)), and the SCAD penalty (Fan and Li (2001)). The lasso has been extensively used in applied problems, and its statistical scope and limitations are well understood (e.g., Bühlmann and van de Geer (2011) and the references cited therein). A common assumption is \textit{sparsity}, i.e., only a small number of covariates influence the outcome. Several refinements have been proposed, in particular the adaptive lasso (Zou (2006)), which relaxes the rather strict conditions required for consistent covariate selection by the standard lasso.

Measurement error in the covariates is a problem in various high-dimensional data sets. In genomics, examples include gene expression microarray data (Purdom and Holmes (2005), Rocke and Durbin (2001)) and high-throughput sequencing (Benjamini and Speed (2012)). In classical regression models, measurement error is known to cause biased parameter estimates and lack of power (Carroll et al. (2006)). Measurement error in SCAD regression has been studied by, e.g., Liang and Li (2009), Ma and Li (2010), Xu and You (2007). Rosenbaum and Tsybakov (2010) introduced the matrix uncertainty (MU) selector, a modification of the Dantzig selector, which handles measurement error and missing data. Through analytical results for the finite sample case, the MU selector is shown to give good parameter estimation and covariate selection. An improved MU selector is presented in Rosenbaum and Tsybakov (2013). Loh and Wainwright (2012) consider generalized M-estimators with lasso regularization, of which special cases include correction for additive measurement error or missing data. The method is shown to yield estimates close to the true parameters, as measured in the $\ell_{1}$- or $\ell_{2}$-norm, and is computationally feasible in the high-dimensional case, despite its non-convexity. We also mention Chen and Caramanis (2013), who consider high-dimensional measurement error problems with independent covariates, and develop a modified orthogonal matching pursuit algorithm yielding correct sparsity recovery with high probability, also when estimates of the measurement error do not exist.

Since the standard lasso is widely used despite the presence of measurement error, e.g., in genomics data, it is of interest to study which impact measurement error has on the analysis. In the first half of this paper, we thus ask: \emph{Under which conditions can the standard lasso (naive approach) be safely used, and when are correction methods required?} For a linear model with additive measurement error, we demonstrate how measurement error affects estimation and prediction error. In the fixed $p$, large $n$ setting, we also show that the naive lasso yields asymptotically sign consistent covariate selection only under very stringent conditions on the noise. Next, a correction of the lasso loss function for linear models is considered, which compensates for measurement error in the covariates. The estimation error of this correction has been studied earlier by Loh and Wainwright (2012). Here, we derive finite sample conditions under which this corrected lasso yields sign consistent covariate selection, and show that it performs asymptotically as well as the lasso without measurement error. We then go on to consider the lasso for generalized linear models (GLMs), and suggest ways to correct for additive measurement error in GLMs, using the conditional score method by Stefanski and Carroll (1987) together with an efficient projection algorithm by Duchi et al. (2008). Finally, the analytical results for linear regression are illustrated through simulations, and the statistical and computational properties of the suggested correction method for GLMs are studied in simulation experiments. We also illustrate the use of measurement error correction in the lasso for logistic regression in an example with micrarray data. Proofs and additional conditions are given in the Supplementary Material.

\par

\setcounter{chapter}{2}
\setcounter{equation}{0} 
\noindent {\bf 2. Model Setup}

In Sections 3 and 4, we consider a linear regression model with additive measurement error,
\begin{align} \label{eq:MEModel}
\mathbf{y} = \mathbf{X}\boldsymbol{\beta}^{0} + \boldsymbol{\epsilon} \qquad \text{and}\qquad
\mathbf{W} = \mathbf{X} + \mathbf{U},
\end{align}
with observations of $p$ covariates and a continuous response $\mathbf{y} \in \mathbb{R}^{n}$ on $n$ individuals. The true covariates $\mathbf{X}$ are not observed, and instead we have noisy measurements $\mathbf{W}$. The matrix of measurement errors $\mathbf{U} \in \mathbb{R}^{n\times p}$ is assumed to have normally distributed rows, with mean zero and covariance $\boldsymbol{\Sigma}_{uu}$. The model errors $\boldsymbol{\epsilon} = \left(\epsilon_{1},\dots,\epsilon_{n}\right)'$ are i.i.d. normally distributed with mean zero and variance $\sigma^{2}$. In Section 5, the linear model on the left-hand side of (\ref{eq:MEModel}) will be replaced by GLMs, but additive measurement error will still be assumed. 

Let $S_{0} = \{j:\beta_{j}^{0} \neq 0\}$ be the index set of non-zero components of the true coefficient vector $\boldsymbol{\beta}^{0} \in \mathbb{R}^{p}$, and denote the number of relevant covariates by $s_{0} = \text{card}\{S_{0} \}$. Under the sparsity assumption, most components of $\boldsymbol{\beta}^{0}$ are zero, such that $s_{0} << p$. Direct use of error-prone measurements is referred to as the \emph{naive approach} in the measurement error literature, and the naive lasso for a linear model takes the form
\begin{equation}\label{eq:Naivelasso}
\hat{\boldsymbol{\beta}}(\lambda) = \text{arg}~\underset{\boldsymbol{\beta}}{\text{min}}\left\{(1/n)\|\mathbf{y} - \mathbf{W}\boldsymbol{\beta}  \|_{2}^{2} + \lambda \|\boldsymbol{\beta} \|_{1}  \right\},
\end{equation}
where $\lambda>0$ is a regularization parameter. For any $\lambda \geq 0$, define the \emph{active set} of the lasso, $\hat{S}(\lambda) = \{j: \hat{\beta}_{j}(\lambda) \neq 0 \}$. Given $\boldsymbol{\beta}^{0}$, we order the covariates such that $S_{0} = \{1,\dots,s_{0}\}$ and $S_{0}^{c} = \{s_{0}+1,\dots,p\}$. Furthermore, we introduce the partitioning $\mathbf{W} = (\mathbf{W}_{S_{0}}, \mathbf{W}_{S_{0}^{c}})$, where $\mathbf{W}_{S_{0}} \in \mathbb{R}^{n \times s_{0}}$ contains the $n$ measurements of the $s_{0}$ relevant covariates, and $\mathbf{W}_{S_{0}^{c}} \in \mathbb{R}^{n \times (p-s_{0})}$ contains the $n$ measurements of the $(p-s_{0})$ irrelevant covariates. The same notation is used for $\mathbf{X}$ and $\mathbf{U}$. Sample covariance matrices are denoted by $\mathbf{C}$, and subscripts show which covariates are involved. For example, the empirical covariance of the measurements is given by $\mathbf{C}_{ww} = (1/n)\mathbf{W}'\mathbf{W}$. Using $\mathbf{C}_{ww}$ as an example, we partition the covariance matrices on the form
\begin{equation*}
\mathbf{C}_{ww} = \left(
\begin{array}{cc}
\mathbf{C}_{ww}\left(S_{0},S_{0}\right) & \mathbf{C}_{ww}\left(S_{0},S_{0}^{c}\right) \\
\mathbf{C}_{ww}\left(S_{0}^{c},S_{0}\right) & \mathbf{C}_{ww}\left(S_{0}^{c},S_{0}^{c}\right)
\end{array}
\right),
\end{equation*}
where $\mathbf{C}_{ww}\left(S_{0},S_{0}\right) \in \mathbb{R}^{s_{0} \times s_{0}}$ is the covariance of the $s_{0}$ relevant covariates, $\mathbf{C}_{ww}\left(S_{0}^{c} ,S_{0}^{c}\right) \in \mathbb{R}^{(p-s_{0}) \times (p-s_{0})}$ is the covariance of the $p-s_{0}$ irrelevant covariates, and $\mathbf{S}_{ww}\left(S_{0},S_{0}^{c}\right) = \mathbf{S}_{ww}\left(S_{0}^{c},S_{0}\right)' \in \mathbb{R}^{s_{0} \times (p-s_{0})}$ is the covariance of the relevant covariates with the irrelevant covariates. Population covariance matrices are denoted by $\boldsymbol{\Sigma}$, and indexed by subscripts and superscripts in the same way as described for the sample covariance matrices. The true coefficient vector is written on the form $\boldsymbol{\beta}^{0}= ((\boldsymbol{\beta}_{S_{0}}^{0})', (\boldsymbol{\beta}_{S_{0}^{c}}^{0})')'$, where $\boldsymbol{\beta}_{S_{0}}^{0} \in \mathbb{R}^{s_{0}}$ are the non-zero coefficients and $\boldsymbol{\beta}_{S_{0}^{c}}^{0} \in \mathbb{R}^{(p-s_{0})}$ is a vector of zeros. The lasso estimates are divided according to the same pattern, i.e., $\hat{\boldsymbol{\beta}}= ((\hat{\boldsymbol{\beta}}_{S_{0}})',(\hat{\boldsymbol{\beta}}_{S_{0}^{c}})')'$, where $\hat{\boldsymbol{\beta}}_{S_{0}} \in \mathbb{R}^{s_{0}}$, $\hat{\boldsymbol{\beta}}_{S_{0}^{c}} \in \mathbb{R}^{p-s_{0}}$, and the dependence on $\lambda$ is implicit. Note that the elements of $\hat{\boldsymbol{\beta}}_{S_{0}}$ are not necessarily non-zero, neither are the elements of $\hat{\boldsymbol{\beta}}_{S_{0}^{c}}$ necessarily zero. 

Finally, vectors and matrices are written in boldface, and we use the notation $|\mathbf{v}|\leq|\mathbf{w}|$ for vectors $\mathbf{v},\mathbf{w}\in \mathbb{R}^{p}$ to mean that $|v_{i}|\leq |w_{i}|$ for $i=1,\dots,p$, and equivalently for the other relational operators.
\\
\par

\setcounter{chapter}{3}
\setcounter{equation}{0} 
\noindent {\bf 3. Impact of Ignoring Measurement Error}

Using the error-free case as a reference, we will show in this section how known results for estimation, screening and selection are affected by additive measurement error. 

\noindent {\bf 3.1 Estimation Error} 

In the absence of measurement error, the lasso will be consistent for estimation and prediction under certain conditions. In particular, the design $\mathbf{X}$ must satisfy a compatibility condition, and the noise must satisfy $(2/n)\|\boldsymbol{\epsilon}'\mathbf{X}\|_{\infty} \leq \lambda_{0}$ for some constant $\lambda_{0}$. If the regularization parameter is chosen large enough to rule out the noise, the lasso will be consistent. For $\lambda \geq 2\lambda_{0}$, the bound 
\begin{equation*}
(1/n)\left\|\mathbf{X}\left(\hat{\boldsymbol{\beta}} - \boldsymbol{\beta}^{0}\right)\right\|_{2}^{2} + \lambda \left\|\hat{\boldsymbol{\beta}} - \boldsymbol{\beta}^{0}\right\|_{1} \leq \frac{4 \lambda^{2} s_{0}}{\phi_{0}^{2}}
\end{equation*}
holds, where $\phi_{0}$ is a compatibility constant (Bühlmann and van de Geer (2011, Ch. 6)). Then, e.g., for Gaussian errors, $(2/n)\|\boldsymbol{\epsilon}'\mathbf{X}\|_{\infty} \leq \lambda_{0}$ holds with high probability for $\lambda_{0} \asymp \sqrt{\log p/n}$. Hence, as long as $n\to \infty$ faster than $s_{0} \log{p}$, lasso will be consistent for prediction, and if $n\to \infty$ faster than $s_{0}^{2} \log{p}$, lasso will be consistent for estimation in the $\ell_{1}$-norm.

When the covariates are subject to measurement error, there are two noise terms which need to be bounded: the model error $\boldsymbol{\epsilon}$ and the measurement error $\mathbf{U}$. In order to bound the estimation error, we need a compatibility condition involving the observed covariates.

\noindent{\bf Definition 1.} \textit{The compatibility condition holds for the index set $S_{0}$ if, for some $\phi_{0}>0$ and all $\boldsymbol{\gamma}\in \mathbb{R}^{p}$ such that $\|\boldsymbol{\gamma}_{S_{0}^{c}}\|_{1}\leq 3\|\boldsymbol{\gamma}_{S_{0}}\|_{1}$, it holds that
\begin{equation*}
\left\|\boldsymbol{\gamma}_{S_{0}}\right\|_{1}^{2} \leq \frac{s_{0} \left\|\mathbf{W}\boldsymbol{\gamma}\right\|_{2}^{2}}{n\phi_{0}^{2}}.
\end{equation*}}

We now have the following prediction and estimation bound for the lasso with measurement error.

\noindent{\bf Proposition 1.} \textit{Assume the compatibility condition with constant $\phi_{0}$, and that there exists a constant $\lambda_{0}$ such that
\begin{equation}
(2/n)\left\|\left(\boldsymbol{\epsilon} - \mathbf{U}\boldsymbol{\beta}^{0} \right)' \mathbf{W} \right\|_{\infty} \leq \lambda_{0}.
\label{eq:NoiseBoundME}
\end{equation}
Then, with a regularization parameter $\lambda\geq 2\lambda_{0}$, the following bound holds for the naive lasso:
\begin{equation}
(1/n)\left\|\mathbf{W}\left(\hat{\boldsymbol{\beta}} - \boldsymbol{\beta}^{0}\right)\right\|_{2}^{2} + \lambda \left\|\hat{\boldsymbol{\beta}} - \boldsymbol{\beta}^{0}\right\|_{1} \leq \frac{4 \lambda^{2} s_{0}}{\phi_{0}^{2}}.
\label{eq:EstBoundME}
\end{equation}}

The result (\ref{eq:EstBoundME}) shows us that also in the presence of measurement error, the estimation error of the lasso can be bounded. However, the bound (\ref{eq:NoiseBoundME}) contains a term which is quadratic in the measurement error. By the triangle inequality, the bound (\ref{eq:NoiseBoundME}) is implied by
\begin{equation*}
(2/n)\left\|\boldsymbol{\epsilon}'\mathbf{W}\right\|_{\infty} + (2/n)\left\|\left(\boldsymbol{\beta}^{0}\right)'\mathbf{U}'\mathbf{X}\right\|_{\infty}+ (2/n)\left\|\boldsymbol{\beta}^{0}\right\|_{1}\left\|\mathbf{U}'\mathbf{U}\right\|_{\infty}\leq \lambda_{0}.
\end{equation*}
Hence, if all three terms in the expression above converged to zero, the lasso with measurement error would be consistent. However, the term $\mathbf{U}'\mathbf{U}$ converges to $n\boldsymbol{\Sigma}_{uu}$ as $n\to \infty$. Since $\left\|\boldsymbol{\Sigma}_{uu}\right\|_{\infty} \neq 0$, we do not obtain consistency. Indeed, taking an asymptotic point of view, we have the following result.

\noindent{\bf Proposition 2.} \textit{Assume $\lambda \to 0$ as $n\to \infty$. Then, as $n\to \infty$ with fixed $p$,
\begin{equation*}
\hat{\boldsymbol{\beta}} \overset{p}{\to} \boldsymbol{\Sigma}_{ww}^{-1}\boldsymbol{\Sigma}_{xx} \boldsymbol{\beta}^{0}.
\end{equation*}}
In the absence of measurement error, the lasso estimates converge in probability to $\boldsymbol{\beta}^{0}$ under the same conditions (Knight and Fu (2000)). Hence, with a proper scaling of $\lambda$, the bias induced by additive measurement error is the same as for a multivariate linear model (Carroll et al. (2006)).

\noindent {\bf 3.2 Covariate Selection}

We now consider exact recovery of the sign pattern of $\boldsymbol{\beta}^{0}$, which is an important goal, e.g., in high-throughput genomics. In the absence of measurement error, such \emph{sign consistent covariate selection} requires an irrepresentable condition (IC) (Meinshausen and Bühlmann (2006), Zhao and Yu (2006)). In the presence of measurement error, the IC has a new form:

\noindent{\bf Definition 2.} \textit{The IC with Measurement Error (IC-ME) holds if there exists a constant $\theta \in [0,1)$ such that
\begin{equation*}
\left\|\mathbf{C}_{ww}\left(S_{0}^{c},S_{0}\right) \mathbf{C}_{ww}\left(S_{0},S_{0}\right)^{-1} \text{sign}\left( \boldsymbol{\beta}_{S_{0}}^{0}\right) \right\|_{\infty} \leq \theta.
\end{equation*}}

We refer to Zhao and Yu (2006) for a thorough interpretation of the IC. In the presence of measurement error, we need an additional condition to obtain sign consistent covariate selection with high probability:

\noindent{\bf Definition 3.}
\textit{The Measurement Error Condition (MEC) is satisified if
\begin{equation*}
\boldsymbol{\Sigma}_{ww}\left(S_{0}^{c},S_{0}\right)\boldsymbol{\Sigma}_{ww}\left(S_{0},S_{0}\right)^{-1} \boldsymbol{\Sigma}_{uu}\left(S_{0},S_{0}\right) - \boldsymbol{\Sigma}_{uu}\left(S_{0}^{c},S_{0}\right)  = \mathbf{0}.
\end{equation*}}

Note that the MEC applies to population covariance matrices, whereas the IC-ME applies to sample covariance matrices. As the following result shows, the IC-ME is used to obtain a positive lower bound on the probability of sign consistent covariate selection in the finite sample case. The MEC, together with other conditions, is sufficient to obtain sign consistent selection with probability approaching one in the large sample limit, keeping $p$ fixed. Let
\begin{align*}
&\mathbf{Z}_{1} = \mathbf{C}_{ww}\left(S_{0},S_{0}\right)^{-1} \frac{\mathbf{W}_{S_{0}}'}{\sqrt{n}} \\
&\mathbf{Z}_{2} = \sqrt{n} \mathbf{C}_{ww}\left(S_{0},S_{0}\right)^{-1} \mathbf{C}_{wu}\left(S_{0},S_{0}\right) \\
&\mathbf{Z}_{3} = \mathbf{C}_{ww}\left(S_{0}^{c},S_{0}\right) \mathbf{C}_{ww}\left(S_{0},S_{0}\right)^{-1} \frac{\mathbf{W}_{S_{0}}'}{\sqrt{n}} - \frac{\mathbf{W}_{S_{0}^{c}}'}{\sqrt{n}} \\
&\mathbf{Z}_{4} = \sqrt{n}\left(\mathbf{C}_{ww}\left(S_{0}^{c},S_{0}\right) \mathbf{C}_{ww}\left(S_{0},S_{0}\right)^{-1} \mathbf{C}_{wu}\left(S_{0},S_{0}\right) - \mathbf{C}_{wu}\left(S_{0}^{c},S_{0}\right)\right).
\end{align*}
We now have the following result for covariate selection with the lasso in the presence of measurement error:

\noindent{\bf Theorem 1.}\textit{
Assume the IC-ME holds with constant $\theta$. Then
\begin{equation*}
P\left(\text{sign}\left(\hat{\boldsymbol{\beta}}\right) = \text{sign}\left(\boldsymbol{\beta}^{0}\right)\right) \geq P\left(A \cap B\right),
\end{equation*}
for the events
\begin{align*}
A &= \left\{ \left|\mathbf{Z}_{1} \boldsymbol{\epsilon} -\mathbf{Z}_{2} \boldsymbol{\beta}_{S_{0}}^{0} \right|  
< \sqrt{n} \left(\left|\boldsymbol{\beta}_{S_{0}}^{0} \right|  - \frac{\lambda}{2}\left|\mathbf{C}_{ww}\left(S_{0},S_{0}\right)^{-1} \text{sign}\left(\boldsymbol{\beta}_{S_{0}}^{0}\right)\right| \right)\right\}
\end{align*}
and
\begin{align*}
B &= \bigg\{ \left|\mathbf{Z}_{3} \boldsymbol{\epsilon} - \mathbf{Z}_{4}\boldsymbol{\beta}_{S_{0}}^{0} \right|  < \frac{\lambda \sqrt{n}}{2 } \left(1-\theta\right) \mathbf{1}\bigg\}.
\end{align*}
If, in addition, the MEC is satisfied and $|\boldsymbol{\beta}_{S_{0}}^{0}| > |\boldsymbol{\Sigma}_{ww}\left(S_{0},S_{0} \right)^{-1} \boldsymbol{\Sigma}_{uu}\left(S_{0},S_{0} \right) \boldsymbol{\beta}_{S_{0}}^{0}|$, then
\begin{equation*}
P\left(\text{sign}(\hat{\boldsymbol{\beta}}) = \text{sign}(\boldsymbol{\beta}^{0})\right) = 1 -o(\exp(-n^{c})), \text{ for some } c\in [0,1),
\end{equation*}
if $\lambda\to 0$ and $\lambda n^{(1-c)/2}\to \infty$ as $n\to \infty$ with fixed $p$.}

Event $A$ implies that the relevant covariates are estimated with correct sign. Given $A$, event $B$ implies that the coefficients of the irrelevant covariates are correctly set to zero. As in the case with perfectly measured covariates (Zhao and Yu (2006)), the left-hand sides of $A$ and $B$ involve the model error $\boldsymbol{\epsilon}$, which needs to be bounded. Due to the presence of measurement error, there are also terms involving $\boldsymbol{\beta}_{S_{0}}^{0}$ on the left-hand side in both events. That is, due to measurement error, terms involving products of the measurement error with the true covariates affect the covariate selection performance. The same dependence on $\boldsymbol{\beta}^{0}$ is seen in the results of Chen and Caramanis (2013) for covariate selection by orthogonal matching pursuit. The events $A$ and $B$ also illustrate the trade-off between choosing $\lambda$ small enough to include the relevant covariates (increasing $P(A)$) and large enough to discard the irrelevant covariates (increasing $P(B)$).

The MEC holds when $\boldsymbol{\Sigma}_{xx}(S_{0}^{c},S_{0}) = \boldsymbol{\Sigma}_{uu}(S_{0}^{c},S_{0}) = \mathbf{0}$, ensuring sign consistent covariate selection with high probability in the large sample limit. Chen and Caramanis (2013, Th. 3) prove that the orthogonal matching pursuit algorithm with covariates subject to measurement error, identifies the relevant covariates with high probability provided the elements of $\left|\boldsymbol{\beta}^{0}_{S_{0}}\right|$ exceed a certain threshold and the covariates are uncorrelated. We have hence shown that a similar result holds for the naive lasso as well. No correlation between the relevant and the irrelevant covariates is an unlikely situation in most applications, unfortunately. The MEC also holds whenever the population covariance matrix of the measurement error has the same form as the population covariance matrix of the true covariates, i.e., $\boldsymbol{\Sigma}_{uu} = c \boldsymbol{\Sigma}_{xx}$ for some constant $c$.

Theorem 1 states that the naive lasso is asymptotically sign consistent if the MEC holds, but this does of course not imply that the lasso is \emph{not} asymptotically sign consistent if the MEC does \emph{not} hold. Useful insight into necessary and sufficient conditions for sign consistent covariate selection can be obtained by considering the case of no model error, $\boldsymbol{\epsilon}=\mathbf{0}$. In the absence of measurement error, the IC is known to be a sharp condition when $\boldsymbol{\epsilon}=\mathbf{0}$: for a finite sample, the lasso will estimate the signs correctly if and only if a version of the IC holds (Bühlmann and van de Geer (2011, Ch. 7)). Our next result states necessary and sufficient conditions for sign consistent covariate selection when $\boldsymbol{\epsilon}=\mathbf{0}$ and the covariates are subject to measurement error. We use the shorthand
\begin{equation*}
\mathbf{Z}_{5} = \mathbf{C}_{ww}\left(S_{0}^{c},S_{0} \right)\mathbf{C}_{ww}\left(S_{0},S_{0} \right)^{-1} \mathbf{C}_{wu}\left(S_{0},S_{0} \right) - \mathbf{C}_{wu}\left(S_{0}^{c},S_{0} \right).
\end{equation*}

\noindent{\bf Proposition 3.}\textit{
Consider the naive lasso in the case of no model error, $\boldsymbol{\epsilon}=\mathbf{0}$. Define the set of detectable covariates by
\begin{align}\label{eq:Prop2BetaMin}
S_{0}^{\text{det}} = &\bigg\{j: \left|\beta_{j}^{0} \right| > \frac{\lambda}{2} \left( \underset{\left\|\boldsymbol{\tau}_{S_{0}}\right\|_{\infty} \leq 1}{\text{sup}} \left\| \mathbf{C}_{ww}\left(S_{0},S_{0} \right)^{-1}\boldsymbol{\tau}_{S_{0}}\right\|_{\infty}\right)   +\left|v_{j} \right|\bigg\},
\end{align}
where 
\begin{equation*}
\mathbf{v} = \left(v_{1},\dots,v_{p}\right)' = \mathbf{C}_{ww}\left(S_{0},S_{0} \right)^{-1} \mathbf{C}_{wu}\left(S_{0},S_{0} \right) \boldsymbol{\beta}_{S_{0}}^{0} .
\end{equation*}
If the IC-ME is satisfied and $\mathbf{Z}_{5}\boldsymbol{\beta}_{S_{0}}^{0} = \mathbf{0}$, then $S_{0}^{\text{det}} \subseteq \hat{S}\left(\lambda\right) \subseteq S_{0}$. Conversely, if $\hat{S}\left(\lambda\right) = S_{0} = S_{0}^{\text{det}}$, then
\begin{align}
&\left\|\mathbf{C}_{ww}\left(S_{0}^{c},S_{0} \right)\mathbf{C}_{ww}\left(S_{0},S_{0} \right)^{-1} \text{sign}\left(\boldsymbol{\beta}_{S_{0}}^{0}\right)  + \frac{2}{\lambda} \mathbf{Z}_{5}\boldsymbol{\beta}_{S_{0}}^{0}  \right\|_{\infty} \leq 1.
\end{align}}

In (\ref{eq:Prop2BetaMin}), the first term is the same as we would have in the absence of measurement error, except that it involves $\mathbf{C}_{ww}$ rather than $\mathbf{C}_{xx}$. The second term in (\ref{eq:Prop2BetaMin}), however, involves the measurement errors and $\boldsymbol{\beta}_{S_{0}}^{0}$. Due to this second term, the lasso cannot detect arbitrarily small coefficients in the presence of measurement error.

\setcounter{chapter}{4}
\setcounter{equation}{0} 
\noindent {\bf 4. Correction for Measurement Error in Lasso: Linear Case}

Sign consistent covariate selection with the naive lasso requires that the MEC is satisfied; a much stronger condition than the IC, which is necessary in the absence of measurement error. Correction for measurement error is thus needed, and in this section we will consider a corrected lasso, which yields sign consistent covariate selection under an IC-type condition. The correction we will use, is motivated by the fact that the loss function of the naive lasso is biased:
\begin{equation*}
E\left(\|\mathbf{y} - \mathbf{W}\boldsymbol{\beta} \|_{2}^{2}\big|~\mathbf{X},\mathbf{y} \right) = \|\mathbf{y} - \mathbf{X}\boldsymbol{\beta} \|_{2}^{2} + n\boldsymbol{\beta}' \boldsymbol{\Sigma}_{uu} \boldsymbol{\beta}.
\end{equation*}
This suggests the definition of the regularized corrected lasso (RCL),
\begin{equation}\label{eq:Correctedlasso}
\hat{\boldsymbol{\beta}}_{\text{RCL}} = \underset{\boldsymbol{\beta} : \|\boldsymbol{\beta}\|_{1} \leq R}{\text{arg~min}}\left\{(1/n) \|\mathbf{y} - \mathbf{W}\boldsymbol{\beta} \|_{2}^{2} - \boldsymbol{\beta}' \boldsymbol{\Sigma}_{uu} \boldsymbol{\beta} + \lambda \| \boldsymbol{\beta}\|_{1} \right\},
\end{equation}
introduced by Loh and Wainwright (2012). The loss function of the RCL is always non-convex when $p>n$, and its parameter space must be restricted to the $\ell_{1}$-ball $\{\boldsymbol{\beta}:\|\boldsymbol{\beta}\|_{1} \leq R\}$ with some finite radius $R$ to avoid trivial solutions. There are thus two regularization parameters, $\lambda$ and $R$. A related problem is the constrained corrected lasso (CCL),
\begin{equation}\label{eq:ConstrainedCorrectedlasso}
\hat{\boldsymbol{\beta}}_{\text{CCL}} = \underset{\boldsymbol{\beta} : \|\boldsymbol{\beta}\|_{1} \leq \kappa}{\text{arg~min}}\left\{ (1/n)\|\mathbf{y} - \mathbf{W}\boldsymbol{\beta} \|_{2}^{2} - \boldsymbol{\beta}' \boldsymbol{\Sigma}_{uu} \boldsymbol{\beta} \right\},
\end{equation}
where $\kappa$ is the only constraint parameter, to be chosen by some model selection procedure. Unless distinction is necessary, we will refer to both as the corrected lasso. The same correction has been proposed for linear regression with the SCAD penalty (Liang and Li (2009)). Since the lasso does not possess the oracle property of the SCAD (Fan and Li (2001)), the results of those papers do not immediately hold for the lasso. The corrected lasso has already been shown to yield good estimation bounds (Loh and Wainwright (2012)), and we will now study its capacity for sign consistent selection. We first define an IC for the corrected lasso:

\noindent{\bf Definition 4.}\textit{
The Irrepresentable Condition for the Corrected lasso (IC-CL) holds if the matrix $\mathbf{C}_{ww}(S_{0},S_{0})- \boldsymbol{\Sigma}_{uu}(S_{0},S_{0})$ is invertible, and there exists a constant $\theta \in [0,1)$ such that
\begin{equation*}
\left\|\left(\mathbf{C}_{ww}(S_{0}^{c},S_{0})- \boldsymbol{\Sigma}_{uu}(S_{0}^{c},S_{0})\right) \left(\mathbf{C}_{ww}(S_{0},S_{0})- \boldsymbol{\Sigma}_{uu}(S_{0},S_{0})\right)^{-1} \text{sign}\left( \boldsymbol{\beta}_{S_{0}}^{0}\right) \right\|_{\infty} \leq \theta.
\end{equation*}}

When the empirical covariance matrices are replaced by population covariance matrices, the IC-CL reduces to the standard IC without measurement error. Before stating the main result of the section, we introduce the shorthands
\begin{align*}
\mathbf{Z}_{6} &= \left(\mathbf{C}_{ww}\left(S_{0},S_{0}\right) -\boldsymbol{\Sigma}_{uu}\left(S_{0},S_{0}\right) \right)^{-1} \frac{\mathbf{W}_{S_{0}}'}{\sqrt{n}} \\
\mathbf{Z}_{7} &= \sqrt{n} \left(\mathbf{C}_{ww}\left(S_{0},S_{0}\right) - \boldsymbol{\Sigma}_{uu}\left(S_{0},S_{0}\right)\right)^{-1} \left(\mathbf{C}_{wu}\left(S_{0},S_{0}\right) - \boldsymbol{\Sigma}_{uu}\left(S_{0},S_{0}\right)\right) \\
\mathbf{Z}_{8} &= \left(\mathbf{C}_{ww}\left(S_{0}^{c},S_{0}\right)-\boldsymbol{\Sigma}_{uu}\left(S_{0},S_{0}\right)\right) \left( \mathbf{C}_{ww}\left(S_{0},S_{0}\right)- \boldsymbol{\Sigma}_{uu}\left(S_{0},S_{0}\right)\right)^{-1} \frac{\mathbf{W}_{S_{0}}'}{\sqrt{n}} \\ 
&\qquad  -\frac{\mathbf{W}_{S_{0}^{c}}'}{\sqrt{n}} \\
\mathbf{Z}_{8} &= \sqrt{n}\big(\left(\mathbf{C}_{ww}\left(S_{0}^{c},S_{0}\right) - \boldsymbol{\Sigma}_{uu}\left(S_{0}^{c},S_{0}\right)\right) \left(\mathbf{C}_{ww}\left(S_{0},S_{0}\right)- \boldsymbol{\Sigma}_{uu}\left(S_{0},S_{0}\right)\right)^{-1}\\
&\qquad  \left(\mathbf{C}_{wu}\left(S_{0},S_{0}\right)- \boldsymbol{\Sigma}_{uu}\left(S_{0},S_{0}\right)\right) - \left(\mathbf{C}_{wu}\left(S_{0}^{c},S_{0}\right) - \boldsymbol{\Sigma}_{uu}\left(S_{0}^{c},S_{0}\right)\right)\big).
\end{align*}
We now have the following result for covariate selection with the corrected lasso in the presence of measurement error.

\noindent{\bf Theorem 2.}\textit{
Assume the IC-CL holds with constant $\theta$. Let $\hat{\boldsymbol{\beta}}$ denote a local optimum of the RCL. If $\hat{\boldsymbol{\beta}}$ lies in the interior of the feasible set, i.e., $\|\hat{\boldsymbol{\beta}}\|_{1} < R$, then
\begin{equation}\label{eq:ProbLowerBound}
P\left(\text{sign}(\hat{\boldsymbol{\beta}}) = \text{sign}(\boldsymbol{\beta}^{0})\right) \geq P(A \cap B),
\end{equation}
for the events
\begin{align}\label{eq:MeasErrCondACL}
A &= \bigg\{ \left|\mathbf{Z}_{6} \boldsymbol{\epsilon} - \mathbf{Z}_{7} \boldsymbol{\beta}_{S_{0}}^{0}\right|   <\\ \nonumber
&\qquad \sqrt{n} \left(\left|\boldsymbol{\beta}_{S_{0}}^{0} \right|  - \frac{\lambda}{2}\left|\left(\mathbf{C}_{ww}\left(S_{0},S_{0}\right) -\boldsymbol{\Sigma}_{uu}\left(S_{0},S_{0}\right)\right)^{-1} \text{sign}\left(\boldsymbol{\beta}_{S_{0}}^{0}\right)\right| \right)\bigg\}
\end{align}
and
\begin{align}\label{eq:ConditionBCL}
B &= \bigg\{ \left|\mathbf{Z}_{8}\boldsymbol{\epsilon} -
 \mathbf{Z}_{9}\boldsymbol{\beta}_{S_{0}}^{0} \right|  
< \frac{\lambda\sqrt{n}}{2 } \left(1-\theta\right) \mathbf{1}\bigg\}.
\end{align}
Furthermore, 
\begin{equation}
P\left(\text{sign}(\hat{\boldsymbol{\beta}})= \text{sign}(\boldsymbol{\beta}^{0})\right) = 1 - o(\exp(-n^{c})), \text{ for some } c \in [0,1),
\end{equation}
if $\lambda_{n} \to 0$ and $\lambda_{n}n^{(1-c)/2}\to \infty$ as $n\to \infty$ with fixed $p$.
}

As pointed out by a referee, the condition $\|\hat{\boldsymbol{\beta}}\|_{1} < R$ is required, since the KKT conditions do not characterize critical points on the boundary of the feasible set. A local optimum at the boundary of the feasible set, $\|\hat{\boldsymbol{\beta}}\|_{1} = R$, may arise if the loss function
\begin{equation*}
(1/n)\left\|\mathbf{y} - \mathbf{W}\boldsymbol{\beta} \right\|_{2}^{2} - \boldsymbol{\beta}' \boldsymbol{\Sigma}_{uu} \boldsymbol{\beta} + \lambda \left\|\boldsymbol{\beta}\right\|_{1}
\end{equation*}
is decreasing when going from a $\boldsymbol{\beta}$ just inside the feasible set to a $\boldsymbol{\beta}$ just outside the feasible set. However, a consequence of Theorem 2 in Loh and Wainwright (2012) is that the distance $\|\hat{\boldsymbol{\beta}} - \boldsymbol{\beta}^{0}\|_{1}$ for any local optimum $\hat{\boldsymbol{\beta}}$ is $O(s_{0} \sqrt{\log(p)/n})$. Hence, for sufficiently large $n$, all local optima are contained in a small $\ell_{1}$-ball around $\boldsymbol{\beta}^{0}$, and we can choose $R$ such that the feasible set contains all these optima.

Our analysis of the naive lasso in Section 3 showed that sign consistent selection in that case required the very strict MEC. The corrected lasso, on the other hand, performs sign consistent selection under the weaker IC-CL, which is very similar to the IC. 

\par

\setcounter{chapter}{5}
\setcounter{equation}{0} 
\noindent {\bf 5. Conditional Scores Lasso for GLMs}

We now consider a generalized linear model (GLM), for which ${Y}$ given $\mathbf{X}$ has density
\begin{equation*}
f(y | \mathbf{x}, \boldsymbol{\Theta}) = \exp\left\{\frac{y \eta - \mathcal{D}(\eta)}{\phi} + c(y,\phi)  \right\},
\end{equation*}
where $\eta = \mu + \mathbf{x}'\boldsymbol{\beta}$, $c(\cdot)$ and $\mathcal{D}(\cdot)$ are functions, and $\boldsymbol{\Theta}=(\mu,\boldsymbol{\beta},\phi)$ is the vector of unknown parameters, where $\mu$ is the intercept and $\phi$ is the dispersion parameter. In the classical case, and in the absence of measurement error, a consistent estimate $\hat{\boldsymbol{\Theta}}$ is obtained by maximum likelihood estimation. When the covariates are subject to additive measurement error, unbiased score functions can be constructed using the conditional scores method of Stefanski and Carroll (1987), yielding consistent estimators of $\boldsymbol{\Theta}$. The method is reviewed by Carroll et al. (2006, Ch. 7), and we will follow their notation. The basic idea is to introduce the sufficient statistic for $\mathbf{x}$,
\begin{equation*}
\boldsymbol{\delta} = \mathbf{w} + y \boldsymbol{\Sigma}_{uu}\boldsymbol{\beta}/\phi,
\end{equation*}
and obtain the conditional density
\begin{align}
f(y | \boldsymbol{\delta}, \boldsymbol{\Theta}, \boldsymbol{\Sigma}_{uu}) =  \exp\left\{\frac{y \eta_{*} - \mathcal{D}_{*}\left(\eta_{*}, \phi, \boldsymbol{\beta}' \boldsymbol{\Sigma}_{uu} \boldsymbol{\beta}\right)}{\phi}  + c_{*}\left(y,\phi, \boldsymbol{\beta}' \boldsymbol{\Sigma}_{uu} \boldsymbol{\beta}\right)\right\},
\label{eq:ModifiedMass}
\end{align}
where $\eta_{*}$, $c_{*}(\cdot)$ and $\mathcal{D}_{*}(\cdot)$ are modifications of the functions used in the absence of measurement error. Assuming the dispersion parameter $\phi$ is known, as it is for logistic and Poisson regression, consistent estimators of $(\mu, \boldsymbol{\beta}')'$ are now obtained by solving the estimating equation 
\begin{equation}\label{eq:ConditionalScore}
\sum_{i=1}^{n}
\begin{array}{c}
( y_{i} - \frac{\partial}{\partial \eta_{*}}\mathcal{D}_{*} ) \left( \begin{array}{c}
1 \\
\boldsymbol{\delta}_{i}
\end{array} 
\right)
\end{array}
 = \mathbf{0}.
\end{equation}

This suggests a possible way of obtaining corrected lasso estimates for GLMs with measurement error, by plugging the estimating equation (\ref{eq:ConditionalScore}) into the projected gradient algorithm used by Loh and Wainwright (2012). In particular, the iteration scheme
\begin{align}\label{eq:InterceptIterate}
\mu^{s+1} &= \mu^{s} + \alpha \sum_{i=1}^{n}\left( y_{i} - \frac{\partial}{\partial \eta_{*}}\mathcal{D}_{*}(\eta_{*i}^{s},(\boldsymbol{\beta}^{s})'\boldsymbol{\Sigma}_{uu}\boldsymbol{\beta}^{s} ) \right) \\
\label{eq:SlopeIterate}
\boldsymbol{\beta}^{s+1} &= \boldsymbol{\Pi}_{\mathcal{B}(\kappa)} \left\{ \boldsymbol{\beta}^{s} + \alpha \sum_{i=1}^{n}\left( y_{i} - \frac{\partial}{\partial \eta_{*}}\mathcal{D}_{*}(\eta_{*i}^{s},(\boldsymbol{\beta}^{s})'\boldsymbol{\Sigma}_{uu}\boldsymbol{\beta}^{s} ) \right)\boldsymbol{\delta}_{i}^{s} \right\},
\end{align}
for $s=1,2,\dots$, until convergence, where $\boldsymbol{\Pi}_{\mathcal{B}(\kappa)} (\cdot)$ denotes projection onto $\mathcal{B}(\kappa)$, $\alpha$ is the stepsize and $\eta_{*i}^{s}$ is the value of $\eta_{*}$ for subject $i$ at iteration $s$, will give regression coefficients constrained to the $\ell_{1}$-ball
\begin{equation*}
\mathcal{B}(\kappa) = \{ \boldsymbol{\beta}\in \mathbb{R}^{p} : \|\boldsymbol{\beta}\|_{1} \leq \kappa\}.
\end{equation*}
The projection $\boldsymbol{\Pi}_{\mathcal{B}(\kappa)} (\cdot)$ can be performed by an efficient algorithm proposed by Duchi et al. (2008). The theoretical results of Loh and Wainwright (2012) do not necessarily apply to GLMs, so we cannot guarantee that the local optima found by iteration (\ref{eq:InterceptIterate})-(\ref{eq:SlopeIterate}) will be close to the global optimum. However, we provide empirical results which suggest that this algorithm is indeed useful.

Selection of the constraint parameter $\kappa$ by standard cross-validation requires a loss function. Hanfelt and Liang (1997) construct an approximate likelihood for this model by path-dependent integration, but a simpler alternative may be to use stability selection (Meinshausen and Bühlmann (2010)) or the `elbow rule' (Rosenbaum and Tsybakov (2010, Fig. 1)), for which no loss function is required. Conditional score functions for GLMs with measurement error can be straightforwardly derived (Carroll et al. (2006, Ch. 7)). We will here consider logistic and Poisson regression, for which $\phi=1$ and corrected lasso estimates are easily obtained. 

\noindent {\bf 5.1 Logistic Regression}

Logistic regression with the lasso penalty has been used, e.g., in detection of differentially expressed genetic markers in case/control studies (Ayers and Cordell (2010), Wu et al. (2009)). Here, we consider binomial logistic regression, with response $y_{i} \sim \text{B}\left(1,H(\eta)\right)$, $i=1,\dots,n$, where $H(\eta) = \{1+\exp(-\eta)\}^{-1}$ is the logit function and $\text{B}(\cdot)$ denotes the binomial distribution. When the covariates are subject to additive measurement error, the terms in the conditional density (\ref{eq:ModifiedMass}) are:
\begin{align*}
&\eta_{*} = \mu + \boldsymbol{\beta}'(\mathbf{w} + y \boldsymbol{\Sigma}_{uu}\boldsymbol{\beta}) \\
&c_{*}(y, \boldsymbol{\beta}'\boldsymbol{\Sigma}_{uu}\boldsymbol{\beta}) = (-y^{2}/2)\boldsymbol{\beta}'\boldsymbol{\Sigma}_{uu}\boldsymbol{\beta} \\
&\mathcal{D}_{*}(\eta_{*},  \boldsymbol{\beta}'\boldsymbol{\Sigma}_{uu}\boldsymbol{\beta}) = \log\left\{1 + \exp\left(\eta_{*} - (1/2) \boldsymbol{\beta}'\boldsymbol{\Sigma}_{uu}\boldsymbol{\beta} \right) \right\},
\end{align*}
and
\begin{align*}
\frac{\partial \mathcal{D}_{*}}{\partial \eta_{*}} = H\left\{\eta_{*} - (1/2) \boldsymbol{\beta}'\boldsymbol{\Sigma}_{uu}\boldsymbol{\beta}\right\}.
\end{align*}
Hence, the iteration scheme (\ref{eq:InterceptIterate})-(\ref{eq:SlopeIterate}) becomes
\begin{align*}
&\mu^{s+1} = \mu^{s} + \alpha \sum_{i=1}^{n}\left( y_{i} - H\left\{ \mu^{s} + (\boldsymbol{\beta}^{s})'\mathbf{w}_{i} + (y_{i}-1/2)(\boldsymbol{\beta}^{s})' \boldsymbol{\Sigma}_{uu} \boldsymbol{\beta}^{s} \right\}\right), \\
&\boldsymbol{\beta}^{s+1} = \\
&\boldsymbol{\Pi}_{\mathcal{B}(\kappa)}\left\{\boldsymbol{\beta}^{s} + \alpha \sum_{i=1}^{n}\left( y_{i} - H\left\{ \mu^{s} +(\boldsymbol{\beta}^{s})'\mathbf{w}_{i} + (y_{i}-1/2)(\boldsymbol{\beta}^{s})' \boldsymbol{\Sigma}_{uu} \boldsymbol{\beta}^{s} \right\}\right)(\mathbf{w}_{i} + y_{i} \boldsymbol{\Sigma}_{uu} \boldsymbol{\beta}^{s}) \right\},
\end{align*}
for $s=1,2,\dots$ until convergence.

\begin{figure}%
\includegraphics[width=\columnwidth]{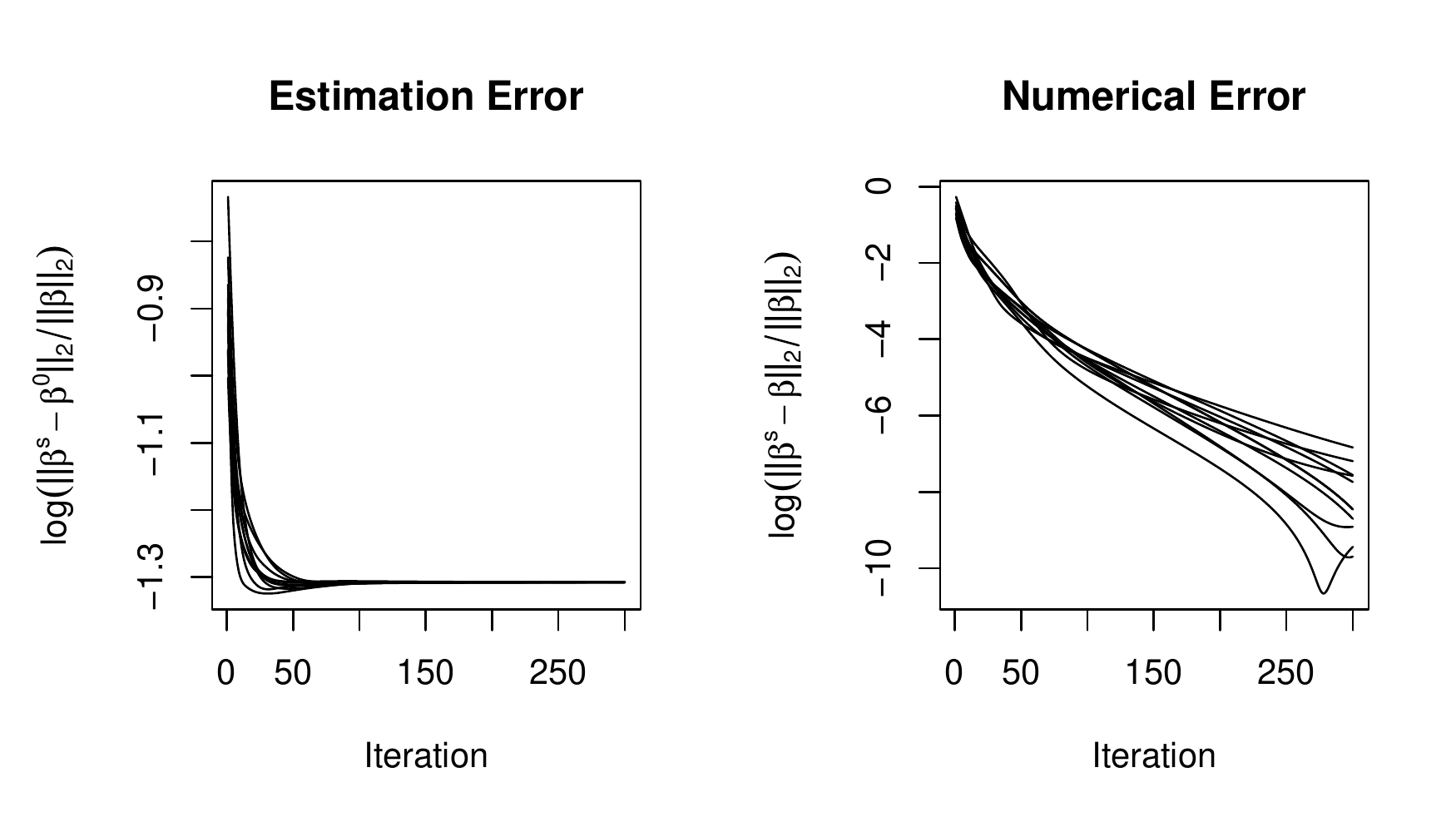}%
\caption{Results of experiments performed in order to assess the convergence properties of the conditional scores iteration scheme for logistic regression, as described in Section 5.1. The left plot shows the logarithm of the $\ell_{2}$ distance to the true regression coefficient $\boldsymbol{\beta}^{0}$ as a function of the iteration count, and the right plot shows the distance to the estimate obtained in the first run, for $10$ consecutive runs with random starting points.}%
\label{fig:LogisticDiagnostics}%
\end{figure}

We performed a simple experiment, similar to Loh and Wainwright (2012, Fig. 2), to assess the convergence properties of this iteration scheme. Setting $n=100$, $p=500$, and $\boldsymbol{\beta}^{0} = (1,1,1,1,1,0,\dots,0)$, we generated a matrix $\mathbf{X} \sim \mathcal{N}(\mathbf{0}, \mathbf{I}_{p})$. A response vector $\mathbf{y}$ with elements $y_{i} \sim B(1,H(\mathbf{x}_{i}'\boldsymbol{\beta}^{0})), ~ i=1,\dots,n$, was then generated, as well as a measurement matrix $\mathbf{W} = \mathbf{X} + \mathbf{U}$, where $\mathbf{U} \sim \mathcal{N}(\mathbf{0}, (0.2)\mathbf{I}_{p})$, i.e., the measurement errors were i.i.d. normally distributed with variance $0.2$. Setting $\kappa = \|\boldsymbol{\beta}^{0}\|_{1}/2$, we ran $300$ iterations with stepsize $\alpha =0.01$, obtaining an estimate of $\boldsymbol{\beta}$. We then repeated this procedure $10$ times, each time with a random initial value $\boldsymbol{\beta}^{1}$. The left plot in Figure \ref{fig:LogisticDiagnostics} shows the logarithm of the relative estimation error in each of the $10$ runs. Starting out at different values, we see that they all converge to a value around $-1.3$. The right plot shows the logarithm of the $\ell_{2}$ distance between each of the $10$ iterates with random starting points, and the estimate obtained in the first run. As we see, this numerical error gets very small as the number of iterations increases, while the estimation error stabilizes. Similar results were obtained for different values of $\kappa$, and for different problem dimensions $n$ and $p$. This suggests that local optima do not pose a problem, at least in this particular setting.

\noindent {\bf 5.2 Poisson Regression}

Poisson regression is used when the outcome can be modeled by a Poisson process, $y_{i} \sim \text{Pois}(e^{\eta})$, $i=1,\dots,n$. An example with high-dimensional data is given by Huang et al. (2010), who define the \emph{spatial lasso}, which is applied with a Poisson regression model to study the distribution of tree species in a geographic area. In the case of additive measurement error, the terms in the modified density (\ref{eq:ModifiedMass}) are:
\begin{align*}
&\eta_{*} = \mu + \boldsymbol{\beta}'(\mathbf{w} + y \boldsymbol{\Sigma}_{uu}\boldsymbol{\beta}) \\
&c_{*}(y, \boldsymbol{\beta}'\boldsymbol{\Sigma}_{uu}\boldsymbol{\beta}) = -\log(y!) - (y^{2}/2)\boldsymbol{\beta}'\boldsymbol{\Sigma}_{uu}\boldsymbol{\beta} \\
&\mathcal{D}_{*}(\eta_{*},  \boldsymbol{\beta}'\boldsymbol{\Sigma}_{uu}\boldsymbol{\beta}) = 
\log\left\{ \sum_{z=0}^{\infty} (z!)^{-1}\exp\left\{ z \eta_{*} - (z^2/2) \boldsymbol{\beta}'\boldsymbol{\Sigma}_{uu}\boldsymbol{\beta} \right\} \right\},
\end{align*}
and
\begin{equation*}
\frac{\partial \mathcal{D}_{*}}{\partial \eta_{*}} = \frac{\sum_{z=0}^{\infty} z (z!)^{-1} \exp\left\{z \eta_{*} - (z^{2}/2) \boldsymbol{\beta}' \boldsymbol{\Sigma}_{uu}\boldsymbol{\beta} \right\}  }{\sum_{z=0}^{\infty} (z!)^{-1}\exp\left\{ z \eta_{*} - (z^2/2) \boldsymbol{\beta}'\boldsymbol{\Sigma}_{uu}\boldsymbol{\beta} \right\} }.
\end{equation*}
Hence, the iteration scheme (\ref{eq:InterceptIterate})-(\ref{eq:SlopeIterate}) for Poisson regression involves numerical approximation of the infinite sums in $\partial \mathcal{D}_{*}/\partial \eta_{*}$, but is otherwise straightforward.

\par

\setcounter{chapter}{6}
\setcounter{equation}{0} 
\noindent {\bf 6. Experiments}

\noindent {\bf 6.1 Linear Regression}

We present the results of simulations comparing the naive lasso (\ref{eq:Naivelasso}) to the corrected lasso for linear models. The constrained version of the corrected lasso (\ref{eq:ConstrainedCorrectedlasso}) was used to avoid dealing with more than one regularization parameters. In all simulations, the number of samples was $n=100$, the number of covariates $p=500$ and the sparsity index was either $s_{0} = 5$ or $s_{0} = 10$. The measurement error covariance $\boldsymbol{\Sigma}_{uu}$ is assumed known, but in Section 6.3 we will illustrate an application where it is estimated. Several different covariance matrices $\boldsymbol{\Sigma}_{xx}$ and $\boldsymbol{\Sigma}_{uu}$ were used, and the overall simulation procedure was as follows:
\begin{itemize}
\item A random set of indices $S_{0} \subset \{1,\dots,p\}$ with cardinality $s_{0}$ was generated. The corresponding nonzero entries of $\boldsymbol{\beta}^{0}$ were then generated by drawing $s_{0}$ i.i.d. values from $\mathcal{N}(0,2^{2})$.
\item The matrix $\mathbf{X} \in \mathbb{R}^{n \times p}$ with rows distributed according to $\mathcal{N}(\mathbf{0}, \boldsymbol{\Sigma}_{xx})$ was generated.
\item The response $\mathbf{y} = \mathbf{X}\boldsymbol{\beta}^{0} + \boldsymbol{\epsilon}$ was sampled with $\boldsymbol{\epsilon}$ i.i.d. drawn from $\mathcal{N}(\mathbf{0}, \sigma^{2}\mathbf{I}_{n})$ with $\sigma=0.1$, and $\mathbf{y}$ had its mean subtracted to avoid estimating the intercept.
\item A measurement matrix $\mathbf{W} = \mathbf{X} + \mathbf{U}$ was generated with the rows of $\mathbf{U}$ i.i.d. distributed according to $\mathcal{N}(\mathbf{0}, \boldsymbol{\Sigma}_{uu})$, and $\mathbf{W}$ had its mean subtracted.
\item The naive lasso estimate $\hat{\boldsymbol{\beta}}_{\text{L}}$ was computed using the R package GLMNET (Friedman et al. (2010)), choosing the regularization level $\lambda$ corresponding to the minimum of the $10$-fold cross-validation curve using the \verb!cv.glmnet! function with default parameters.
\item The corrected lasso estimate $\hat{\boldsymbol{\beta}}_{\text{CL}}$ was computed by $10$-fold cross-validation with $100$ candidate constraint parameters $\kappa$ equally spaced in the range $[10^{-3}R,R]$, where $R = 2\|\hat{\boldsymbol{\beta}}_{\text{L}}\|_{1}$. The final value of $\kappa$ was chosen to minimize the cross-validated loss.
\end{itemize}
The whole procedure was repeated $200$ times for each experiment. Tables \ref{tab:LinearSetting1}-\ref{tab:LinearSetting3} summarize the simulation results. TP (true positives) denotes the number of nonzero covariates which were correctly selected by the procedure and FP (false positives) denotes the number of irrelevant covariates which were selected by the procedure, and the two rightmost columns denote the estimation error as measured in the $\ell_{2}$- and $\ell_{1}$-norm, respectively. All results are averages over the $200$ Monte Carlo simulations, and the numbers in parentheses are the corresponding standard errors.

In our first simulation experiment, all elements of $\mathbf{X}$ and $\mathbf{U}$ were i.i.d. Gaussian, with $\boldsymbol{\Sigma}_{xx} = \mathbf{I}_{p}$ and $\boldsymbol{\Sigma}_{uu} = \sigma_{u}^{2}\mathbf{I}_{p}$, where the measurement error variance $\sigma_{u}^{2}$ was either $0.2$ or $0.4$. For comparison, the standard lasso in the absence of measurement error ($\sigma_{u}^{2}=0.0$) was also computed. Table \ref{tab:LinearSetting1} summarizes the results for both $s_{0}=5$ and $s_{0}=10$.  In all cases shown, the naive lasso makes a very large number of false selections compared to the corrected lasso. The estimation errors of the corrected lasso are also consistently smaller than the naive lasso. Finally, the naive lasso is slightly better than the corrected lasso in detecting the relevant covariates. The lasso without measurement error is also seen to have a higher true positive rate and lower false positive rate than any of the cases with measurement error, indicating that measurement error makes the covariate selection problem considerably harder.

\begin{table}[ht]
\begin{center}
\begin{tabular}{lrllll}
 & & TP & FP & $\|\hat{\boldsymbol{\beta}}-\boldsymbol{\beta}^{0}\|_{2}$ & $\|\hat{\boldsymbol{\beta}}-\boldsymbol{\beta}^{0}\|_{1}$ \\
  \hline
 \multirow{1}{*}{$s_{0}=5$, $\sigma_{\mathbf{U}}^{2} = 0.0$} &Naive  & $4.95 ~(0.02)$ & $5.18 ~(0.52)$ & $0.11 ~(0.00)$ & $0.25 ~(0.01)$ \\ 
   \hline
 \multirow{2}{*}{$s_{0}=5$, $\sigma_{\mathbf{U}}^{2} = 0.2$} &Naive  & $4.18 ~(0.06)$ & $24.82 ~(1.35)$ & $1.56 ~(0.04)$ & $5.03 ~(0.20)$ \\ 
 &Corrected &  $4.13 ~(0.06)$ & $17.57 ~(0.47)$ & $0.98 ~(0.03)$ & $3.23 ~(0.12)$ \\ 
  \hline 
 \multirow{2}{*}{$s_{0}=5$, $\sigma_{\mathbf{U}}^{2} = 0.4$}&Naive &  $3.78 ~(0.07)$ & $23.11 ~(1.37)$ & $2.23 ~(0.06)$ & $6.44 ~(0.21)$ \\ 
 &Corrected&  $3.62 ~(0.06)$ & $11.81 ~(0.32)$ & $1.45 ~(0.04)$ & $4.20 ~(0.14)$ \\ 
  \hline
 \multirow{1}{*}{$s_{0}=10$, $\sigma_{\mathbf{U}}^{2} = 0.0$} &Naive  & $9.82 ~(0.03)$ & $4.82 ~(0.32)$ & $0.20 ~(0.00)$ & $0.64 ~(0.01)$ \\ 
  \hline
   \multirow{2}{*}{$s_{0}=10$, $\sigma_{\mathbf{U}}^{2} = 0.2$}&Naive & $7.46 ~(0.09)$ & $34.27 ~(1.26)$ & $2.80 ~(0.05)$ & $11.67 ~(0.29)$ \\ 
 &Corrected&  $7.16 ~(0.09)$ & $19.77 ~(0.39)$ & $2.05 ~(0.05)$ & $7.91 ~(0.20)$ \\ 
  \hline
     \multirow{2}{*}{$s_{0}=10$, $\sigma_{\mathbf{U}}^{2} = 0.4$}&Naive & $6.35 ~(0.10)$ & $27.98 ~(1.21)$ & $3.80 ~(0.07)$ & $14.29 ~(0.31)$ \\ 
 &Corrected&   $5.76 ~(0.10)$ & $11.97 ~(0.28)$ & $3.01 ~(0.08)$ & $10.28 ~(0.28)$  \\ 
  \hline
\end{tabular}
\end{center}
\caption{Comparison of naive and corrected lasso for linear regression, when $\boldsymbol{\Sigma}_{xx} = \mathbf{I}_{p}$ and $\boldsymbol{\Sigma}_{uu} = \sigma_{u}^{2}\mathbf{I}_{p}$.}
\label{tab:LinearSetting1}
\end{table}

Next, a block diagonal $\boldsymbol{\Sigma}_{xx}$ was considered, with $10$ blocks $\mathbf{B}_{1},\dots,\mathbf{B}_{10} \in \mathbb{R}^{50\times 50}$ along the diagonal, and all other elements equal to zero. Each block had a Toeplitz structure, with $(j,k)$th element given by $\left(\mathbf{B}_{l} \right)_{(j,k)} = 0.8^{|j-k|}$, $l=1,\dots,10$, $j,k = 1,\dots,50$. Hence, $10$ groups of $50$ covariates each were correlated with each other, but not with the covariates outside of the group. The measurement errors were still assumed i.i.d. Gaussian with $\boldsymbol{\Sigma}_{uu}=\sigma_{u}^{2}\mathbf{I}_{p}$, as in the previous setting. This could be a plausible model for gene expression data, with the blocks corresponding to genes within a functional group or pathway having strong correlation (Tai and Pan (2007)), and the measurement error corresponding to noisy measurements. Table \ref{tab:LinearSetting2} shows the simulation results. It is clear that the correlations imposed make this a harder problem. In particular the corrected lasso shows a weaker performance in detecting the relevant covariates, but the naive lasso makes an even larger number of false positive selections.

\begin{table}[ht]
\begin{center}
\begin{tabular}{lrllll}
 & & TP & FP & $\|\hat{\boldsymbol{\beta}}-\boldsymbol{\beta}^{0}\|_{2}$ & $\|\hat{\boldsymbol{\beta}}-\boldsymbol{\beta}^{0}\|_{1}$ \\
  \hline
 \multirow{1}{*}{$s_{0}=5$, $\sigma_{\mathbf{U}}^{2} = 0.0$} &Naive  & $4.95 ~(0.02)$ & $6.45 ~(0.48)$ & $0.13 ~(0.00)$ & $0.31~ (0.01)$ \\ 
   \hline
 \multirow{2}{*}{$s_{0}=5$, $\sigma_{\mathbf{U}}^{2} = 0.2$} &Naive  & $4.03 ~(0.06)$ & $31.58 ~(1.18)$ & $2.22 ~(0.06)$ & $7.41~ (0.23)$ \\ 
 &Corrected &  $3.75 ~(0.06)$ & $15.50 ~(0.40)$ & $1.32 ~(0.05)$ & $4.02 ~(0.17)$ \\ 
  \hline 
 \multirow{2}{*}{$s_{0}=5$, $\sigma_{\mathbf{U}}^{2} = 0.4$}&Naive & $3.82 ~(0.06)$ & $34.56 ~(1.28)$ & $2.79 ~(0.07)$ & $9.29~ (0.28)$ \\ 
 &Corrected&  $3.17 ~(0.07)$ & $10.43 ~(0.26)$ & $1.85 ~(0.07)$ & $5.08 ~(0.19)$ \\ 
   \hline
 \multirow{1}{*}{$s_{0}=10$, $\sigma_{\mathbf{U}}^{2} = 0.0$} &Naive  & $9.73 ~(0.04)$ & $11.13 ~(0.51)$ & $0.27 ~(0.01)$ & $0.92~ (0.02)$ \\ 
  \hline
   \multirow{2}{*}{$s_{0}=10$, $\sigma_{\mathbf{U}}^{2} = 0.2$}&Naive & $7.01 ~(0.09)$ & $41.85 ~(1.18)$ & $3.57 ~(0.06)$ & $15.63 ~(0.32)$ \\ 

 &Corrected&  $6.07 ~(0.11)$ & $17.27 ~(0.34) $& $2.62~ (0.07)$ & $9.62 ~(0.26)$ \\
  \hline
     \multirow{2}{*}{$s_{0}=10$, $\sigma_{\mathbf{U}}^{2} = 0.4$}&Naive & $6.40 ~(0.10)$ & $43.88 ~(1.31)$ & $4.51 ~(0.08)$ & $19.38 ~(0.40)$ \\ 
 &Corrected & $4.38 ~(0.10)$ & $11.38 ~(0.24)$ & $4.00 ~(0.12)$ & $13.43~ (0.40)$ \\ 
  \hline
\end{tabular}
\end{center}
\caption{Comparison of naive and corrected lasso for linear regression, when $\boldsymbol{\Sigma}_{xx}$ is block structured and $\boldsymbol{\Sigma}_{uu} = \sigma_{u}^{2}\mathbf{I}_{p}$.}
\label{tab:LinearSetting2}
\end{table}

Finally, the entries of $\mathbf{X}$ were again i.i.d. Gaussian, $\boldsymbol{\Sigma}_{xx} = \mathbf{I}_{p}$, while the measurement errors had power decay correlations according to the Toeplitz structure, with the $(j,k)$th element of $\boldsymbol{\Sigma}_{uu}$ being $\left(\boldsymbol{\Sigma}_{uu}\right)_{(j,k)} = \rho_{u}^{1+|j-k|}$. The diagonal $\rho_{u} = \sigma_{u}^{2}$ took either the value $0.2$ or $0.4$. Table \ref{tab:LinearSetting3} shows the simulation results. The setting without measurement error is here equivalent to the one in Table \ref{tab:LinearSetting1}, and therefore omitted. In this setting, it is really clear that the corrected lasso performs better than the naive approach. At the cost of a slight reduction in the number of correct selections, the corrected lasso substantially reduces the number of false positive selections. Again, the estimation errors of the corrected lasso are consistently smaller than those of the naive lasso. 

Overall, the reduction in false positive selections when using the corrected lasso compared to the naive lasso, was between $24 \%$ and $74 \%$, which is a substantial improvement.

\begin{table}[ht]
\begin{center}
\begin{tabular}{lrllll}
 & & TP & FP & $\|\hat{\boldsymbol{\beta}}-\boldsymbol{\beta}^{0}\|_{2}$ & $\|\hat{\boldsymbol{\beta}}-\boldsymbol{\beta}^{0}\|_{1}$ \\
   \hline
 \multirow{2}{*}{$s_{0}=5$, $\sigma_{\mathbf{U}}^{2} = 0.2$} &Naive  & $4.19 ~(0.06)$ & $23.86 ~(1.12)$ & $1.63 ~(0.04)$ & $5.09 ~(0.16)$ \\
 &Corrected & $4.13 ~(0.06)$ & $18.02 ~(0.49)$ & $1.04 ~(0.03)$ & $3.53 ~(0.11)$ \\  
  \hline 
 \multirow{2}{*}{$s_{0}=5$, $\sigma_{\mathbf{U}}^{2} = 0.4$}&Naive & $3.79 ~(0.06)$ & $24.37 ~(1.46)$ & $2.30 ~(0.06)$ & $6.88~ (0.25)$ \\ 
 &Corrected& $3.58 ~(0.07)$ & $11.87 ~(0.30)$ & $1.50 ~(0.05)$ & $4.34 ~(0.16)$ \\ 
  \hline
   \multirow{2}{*}{$s_{0}=10$, $\sigma_{\mathbf{U}}^{2} = 0.2$}&Naive & $7.36 ~(0.10)$ & $34.77 ~(1.46)$ & $2.89 ~(0.05)$ & $12.08 ~(0.29)$ \\ 
 &Corrected & $7.16 ~(0.09)$ & $19.98 ~(0.39)$ & $2.11 ~(0.04)$ & $8.16 ~(0.18)$ \\ 
  \hline
     \multirow{2}{*}{$s_{0}=10$, $\sigma_{\mathbf{U}}^{2} = 0.4$}&Naive & $6.49 ~(0.10)$ & $32.38 ~(1.42)$ & $3.80 ~(0.06)$ & $14.91 ~(0.32)$ \\ 
 &Corrected & $5.79 ~(0.09)$ & $11.86 ~(0.25)$ & $2.98 ~(0.07)$ & $10.14 ~(0.25)$ \\ 
  \hline
\end{tabular}
\end{center}
\caption{Comparison of naive and corrected lasso for linear regression, when $\boldsymbol{\Sigma}_{xx} = \mathbf{I}_{p}$ and $\boldsymbol{\Sigma}_{uu}$ is a Toeplitz matrix with elements $(\boldsymbol{\Sigma}_{uu})_{j,k} = \rho_{u}^{1+|j-k|}$, $j,k = 1,\dots,p$, and $\rho_{u} = \sigma_{u}^{2}$.}
\label{tab:LinearSetting3}
\end{table}

\noindent {\bf 6.2 Logistic Regression}
We also performed simulation experiments to investigate the merit of the conditional scores lasso for logistic regression, outlined in Section 5.1. The setup was similar to the one described for linear regression in the last section, except that cross-validation was not performed, due to the lack of a loss function. In the next section, we will show how the regularization paramater $\kappa$ can be set manually using an 'elbow rule', but in these simulations. For now, we compare the naive lasso solution for logistic regression
\begin{equation}
\hat{\boldsymbol{\beta}}_{naive}\left(\kappa\right) = \underset{\boldsymbol{\beta}: \left\|\boldsymbol{\beta}\right\|_{1}\leq \kappa}{\text{arg~min}} \left\{\sum_{i=1}^{n} y_{i} \mathbf{w}_{i}'\boldsymbol{\beta} + \log\left(1 - H\left(\mathbf{w}_{i}'\boldsymbol{\beta} \right) \right)\right\}
\label{eq:NaiveLassoLogistic}
\end{equation}
to the corrected estimate $\hat{\boldsymbol{\beta}}_{corr}(\kappa)$ obtained using the conditional scores algorithm, over a range of candidate $\kappa$ values.

To be specific, we used a sample size $n=100$, $p=500$ covariates, of which either $s_{0}=5$ or $s_{0}=10$ were nonzero. We considered only i.i.d. Gaussian covariates and measurement errors, with $\boldsymbol{\Sigma}_{xx}= \mathbf{I}_{p}$ and $\boldsymbol{\Sigma}_{uu} = (0.2)\mathbf{I}_{p}$, i.e., a measurement error variance $\sigma_{u}^{2}=0.2$. Since both covariance matrices are diagonal, we did not randomize over the indices in $S_{0}$, and simply let $\beta_{j}\neq 0$ for $j=1,\dots,s_{0}$. The matrices $\mathbf{X}$ and $\mathbf{U}$ were drawn from $\mathcal{N}(0,\boldsymbol{\Sigma}_{x})$and $\mathcal{N}(0,\boldsymbol{\Sigma}_{uu})$, respectively. The values of $\boldsymbol{\beta}_{S_{0}}^{0}$ were generated by drawing $s_{0}$ i.i.d. values from $\mathcal{N}(0,5^2)$, and the responses $y_{i}, ~i=1,\dots,n$ were sampled from a binomial distribution with mean $H(\mathbf{x}_{i}'\boldsymbol{\beta}^{0})$. The procedure was repeated $200$ times.

\begin{figure}%
\includegraphics[scale=.4]{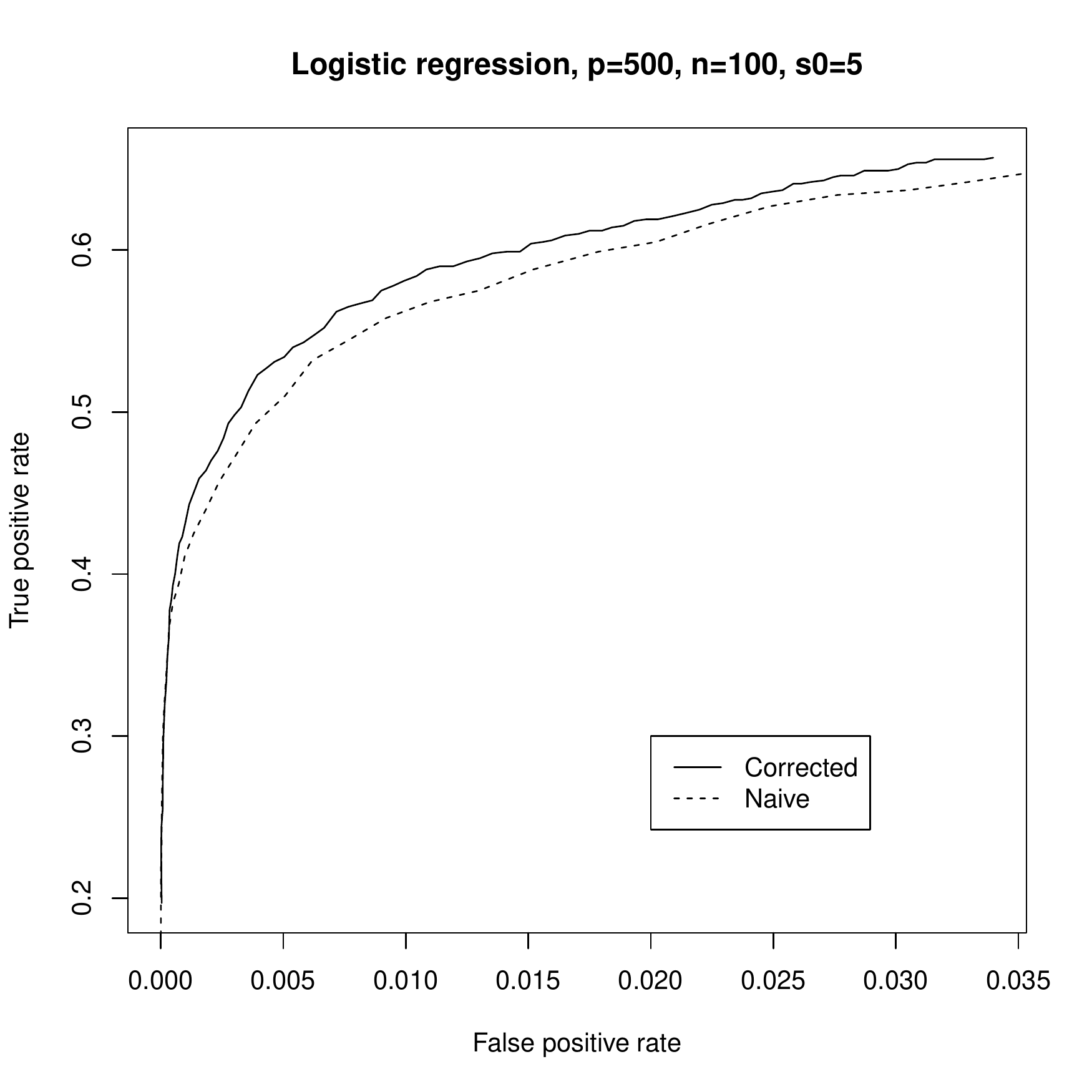}%
\includegraphics[scale=.4]{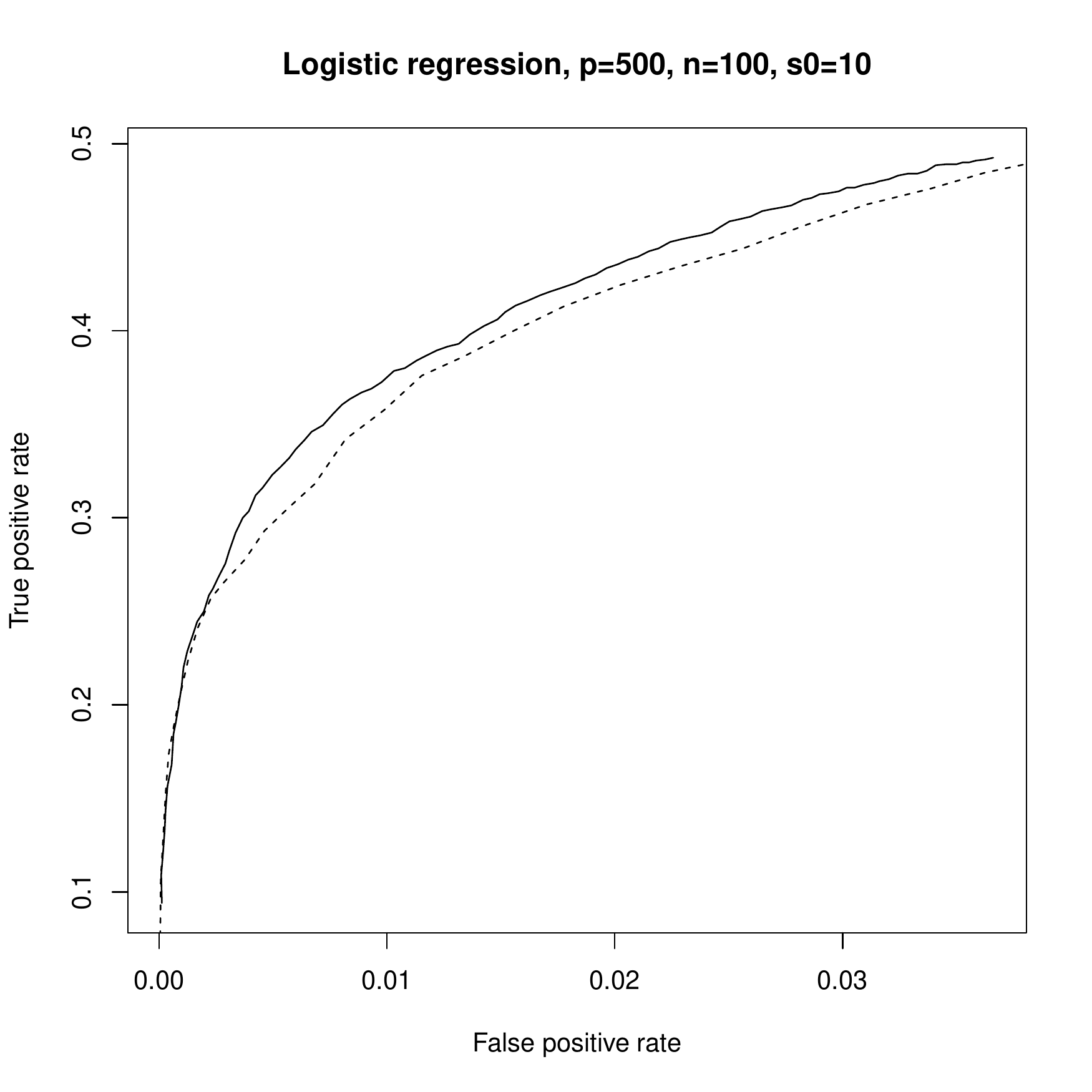}%
\caption{The plots show ROC curves over a range over regularization parameter for the conditional scores lasso ('Corrected') and the standard lasso for logistic regression ('Naive'), for the $s_{0}=5$ case (left) and the $s_{0}=10$ case (right).}%
\label{fig:ROCcurveLogistic}%
\end{figure}

\begin{figure}%
\includegraphics[scale=0.4]{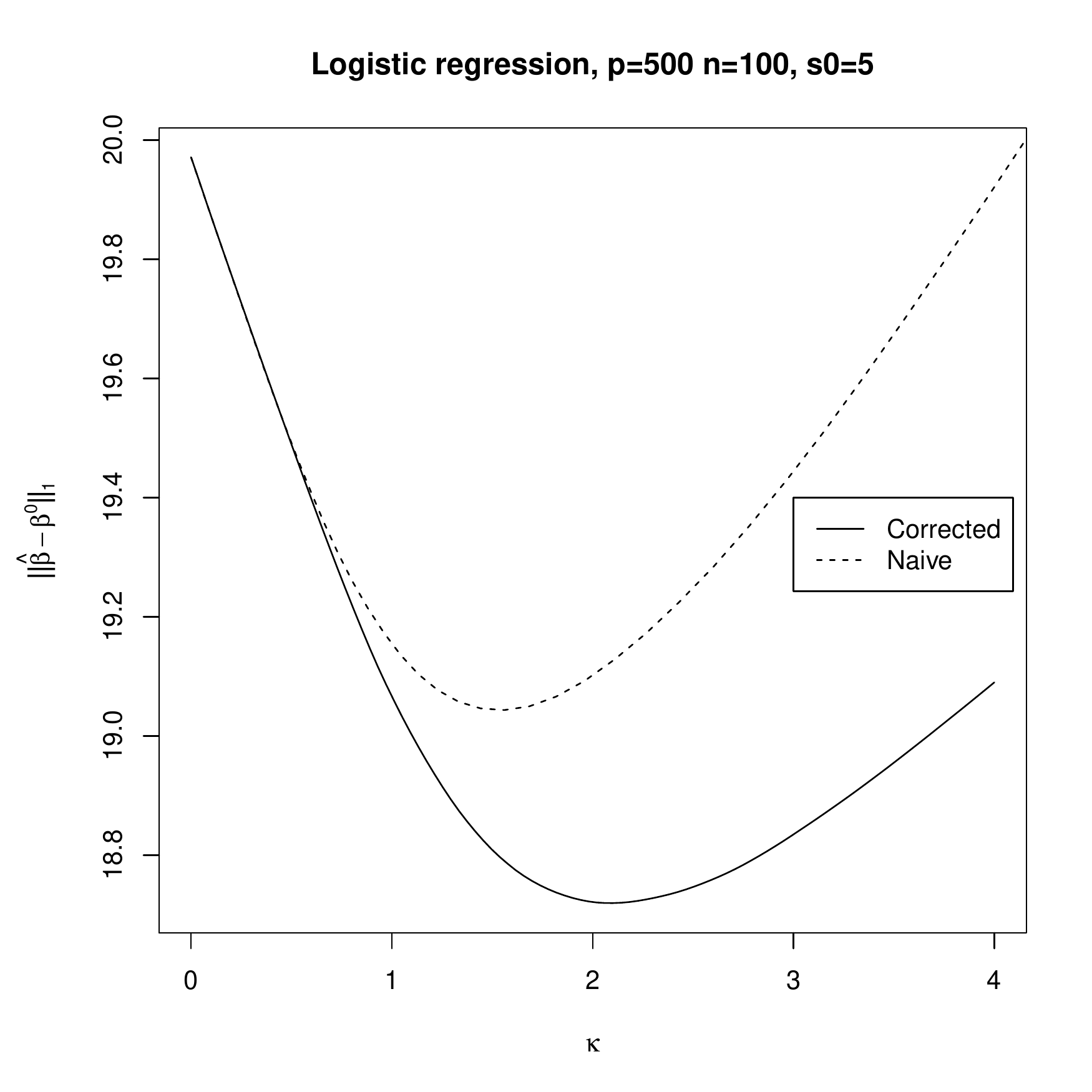}%
\includegraphics[scale=0.4]{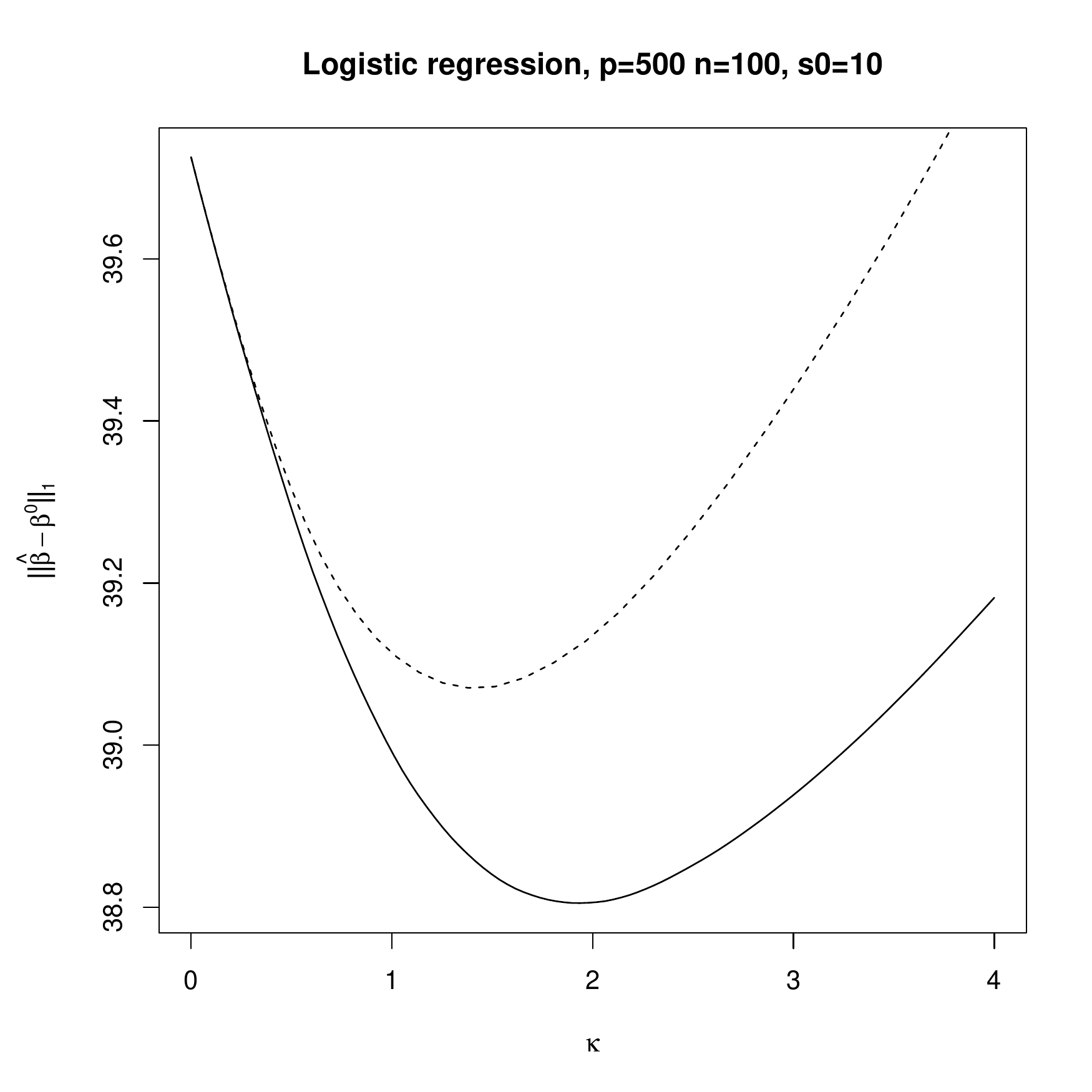}%
\caption{The plots show $\ell_{1}$ estimation error over a range over regularization parameter for the conditional scores lasso ('Corrected') and the standard lasso for logistic regression ('Naive'), for the $s_{0}=5$ case (left) and the $s_{0}=10$ case (right).}%
\label{fig:L1errorLogistic}%
\end{figure}

Figure \ref{fig:ROCcurveLogistic} shows the receiver operating characterics (ROC) curve for the simulations. Particularly in the $s_{0}=5$ case, the conditional scores lasso ('Corrected') is seen to perform better variable selected than the standard lasso for logistic regression ('Naive'). In the $s_{0}=10$ case, the conditional scores lasso is also better, but only marginally. Figure \ref{fig:L1errorLogistic} shows the $\ell_{1}$ estimation error over a range of values of the regularization parameter $\kappa$. Here, the corrected lasso clearly has a lower estimation error than the naive approach in both the $s_{0}=5$ and the $s_{0}=10$ case. These simulations hence suggest that the conditional scores lasso is a useful method for measurement error correction in logistic regression when $p>n$.

\noindent {\bf 6.3 Microarray Data}

We now present an example application of the conditional scores lasso for logistic regression, using an Affymetrix microarray data set publicly available from the ArrayExpress database (www.ebi.ac.uk/arrayexpress) under accession number E-GEOD-10320. The data set contains gene expression measurements of $144$ favorable histology Wilms tumors (FHWT), $53$ of which did relapse (cases) and $91$ of which did not relapse (controls). For the Affymetrix microarrays used, each gene expression is measured by multiple probes. The Bayesian Gene Expression (BGX) Bioconductor package (Hein et al. (2005)) utilizes these replicate measurements to form posterior distributions of the mean gene expression, measured on the log scale, of each gene for each sample. 

In our additive measurement error model, the mean posterior gene expressions $\hat{\mu}_{ij}$ are used as an estimate of the `true' gene expression $x_{ij}$, for $i=1,\dots,n$, $j=1,\dots,p$. Letting $\hat{\boldsymbol{\mu}}_{j} = (\hat{\mu}_{1j},\dots,\hat{\mu}_{nj})'$, $\bar{\mu}_{j} = (1/n)\sum_{i=1}^{n}\hat{\mu}_{ij}$ and $\hat{\sigma}_{j}^{2}=(1/n)\sum_{i=1}^{n}(\hat{\mu}_{ij}-\bar{\mu}_{j})^{2}$, the standardized design matrix $\mathbf{W}$ now has entries
\begin{equation*}
w_{ij} = \frac{\hat{\mu}_{ij} - \bar{\mu}_{j}}{\hat{\sigma}_{j}}, ~i=1,\dots,n, ~j=1,\dots,p.
\end{equation*}
We let $\text{var}(\hat{\mu}_{ij})$ denote the posterior variance of gene expression estimate $\hat{\mu}_{ij}$. Assuming equal measurement error variance across samples, but not across covariates, we estimate the measurement error of gene $j$ by $\hat{\sigma}_{u,j}^{2} = (1/n)\sum_{i=1}^{n}\text{var}(\hat{\mu}_{ij})$, $j=1,\dots,p$. For simplicity, covariance of measurement errors are not considered, so the final estimate $\hat{\boldsymbol{\Sigma}}_{uu}$ on the scale of the standardized $\mathbf{W}$, now has entries
\begin{equation*}
\left( \hat{\boldsymbol{\Sigma}}_{uu}\right)_{j,k} = 
\begin{cases} \hat{\sigma}_{u,j}^{2}/\hat{\sigma}_{j}^{2}, & \mbox{if } j=k \\ 
0, & \mbox{if } j \neq k \end{cases}
\end{equation*}
for $j,k=1,\dots,p$. In many cases, the estimated measurement error variance $\hat{\sigma}_{u,j}^{2}$ is large compared to the between-sample variance of the means $\hat{\sigma}_{j}^{2}$, and for these cases little can be done. We therefore chose to analyze only the $p=1857$ genes for which $\hat{\sigma}_{\mathbf{U},j}^{2} < (1/2) \hat{\sigma}_{j}^{2}$, out of the original $20931$ genes. For the $1857$ selected genes, the naive lasso estimate $\hat{\boldsymbol{\beta}}_{\text{L}}$ was computed by ten-fold cross-validation using GLMNET (Friedman et al. (2010)), yielding $22$ nonzero coefficients. 

\begin{figure}%
\includegraphics[width=8cm]{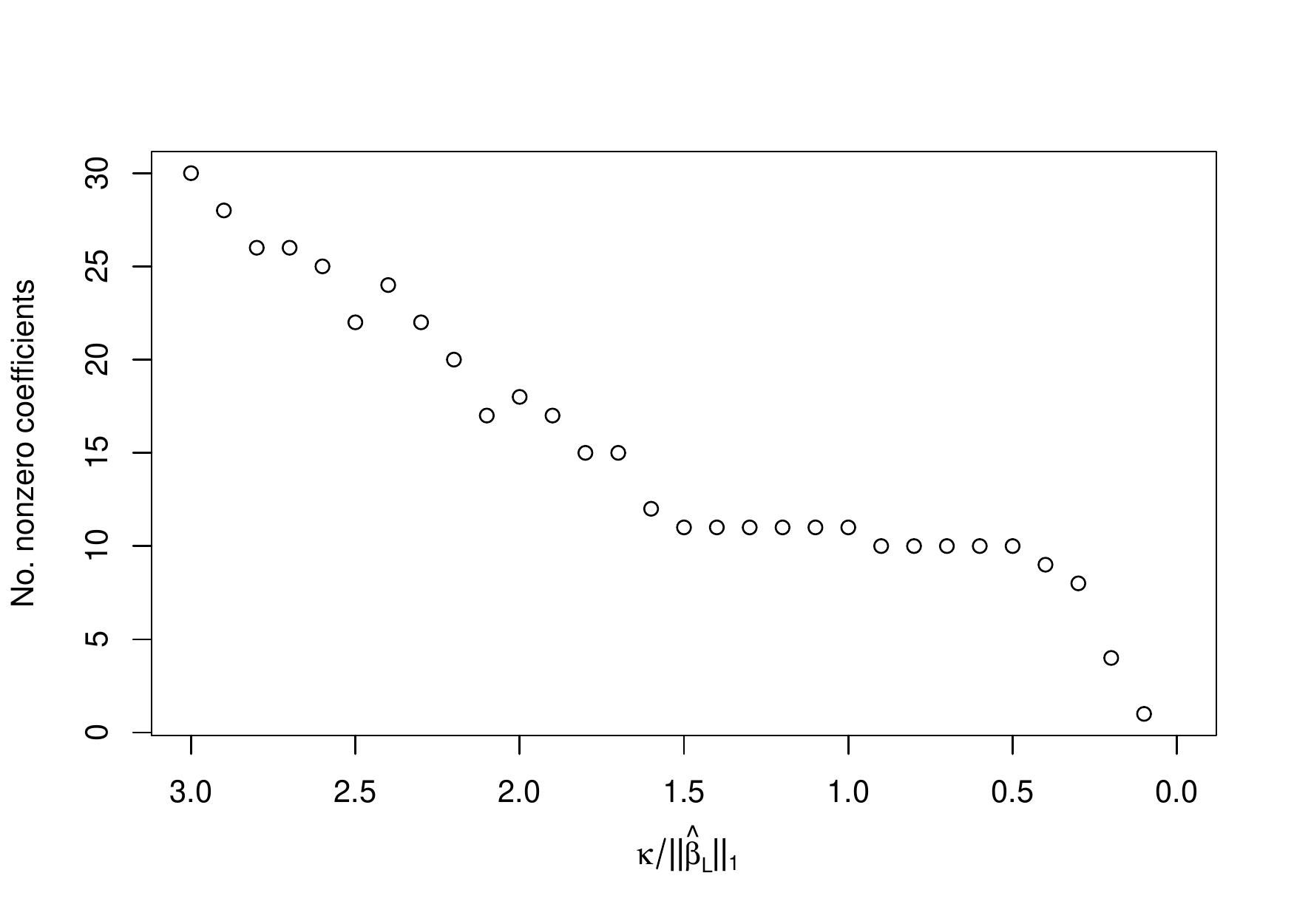}%
\caption{Illustration of the elbow rule for the conditional scores lasso. The number of nonzero coefficients is plotted against the constraint level, here as a fraction of the $\ell_{1}$ norm of the naive lasso estimate.}%
\label{fig:ElbowRule}%
\end{figure}

Since the conditional scores lasso lacks a well-defined loss function, the elbow rule (Rosenbaum and Tsybakov (2010, Fig. 1)) was used to choose the regularization level. The conditional scores solution was computed for a grid of constraint values between $3\|\hat{\boldsymbol{\beta}}_{\text{L}}\|$ and $(0.1)\|\hat{\boldsymbol{\beta}}_{\text{L}}\|$, with spacing $(0.1)\|\hat{\boldsymbol{\beta}}_{\text{L}}\|$. Figure \ref{fig:ElbowRule} shows the number of nonzero coefficient estimates plotted versus the constraint level, and the elbow rule now amounts to selecting $\kappa$ where the curve begins to be flat. The plot in Figure \ref{fig:ElbowRule} is suprisingly good: the number of selected covariates is between $11$ and $10$ for all constraint values between $\kappa=(1.5)\|\hat{\boldsymbol{\beta}}_{\text{L}}\|$ and $\kappa=(0.5)\|\hat{\boldsymbol{\beta}}_{\text{L}}\|$.

\begin{figure}
\includegraphics[width=10cm]{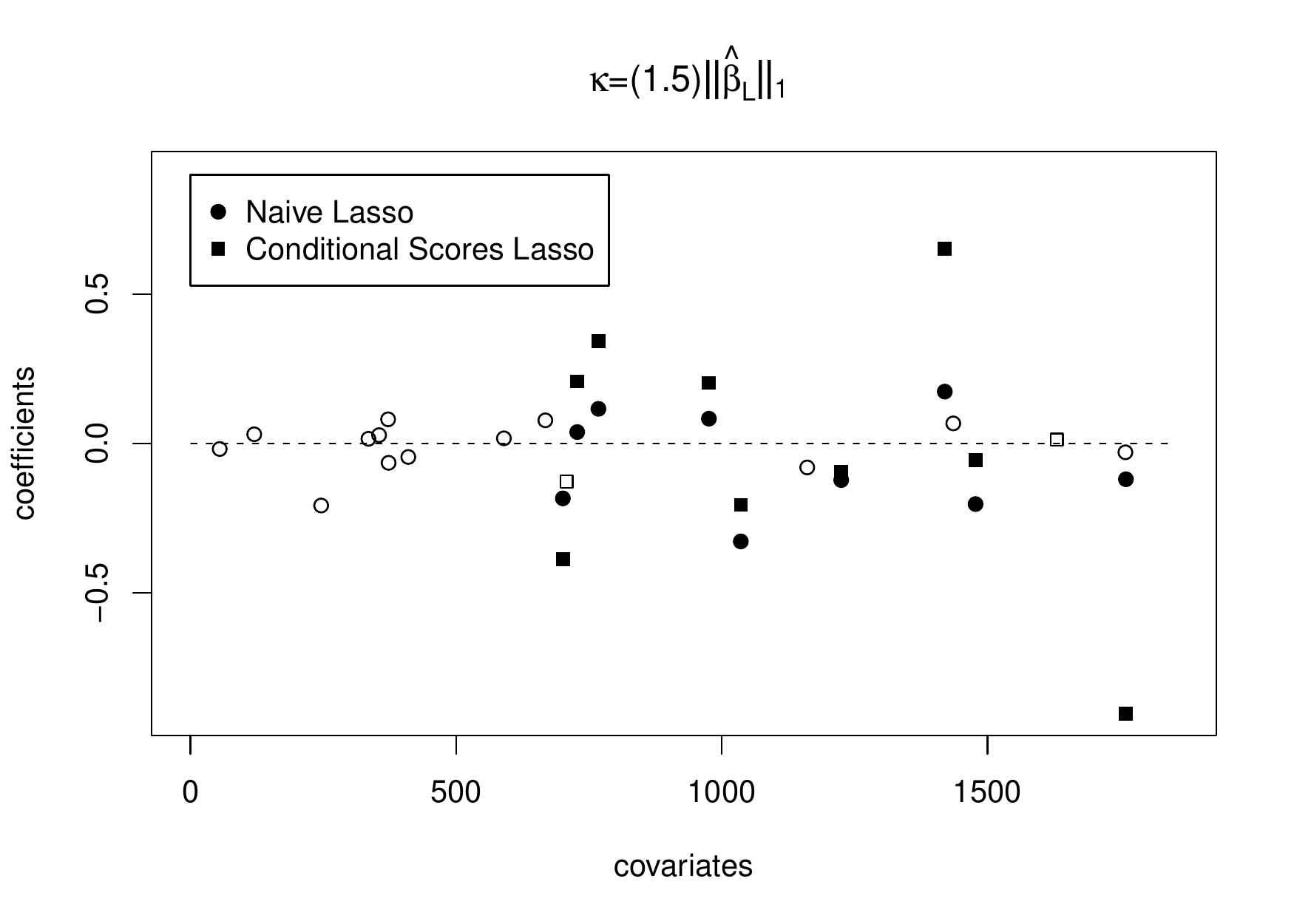}
\caption{Comparison of the coefficient estimates of the $22$ covariates selected by the cross-validated naive lasso and the $11$ covariates selected by the conditional scores lasso. The covariates selected by both methods are plotted with filled circles/squares, whereas those selected by only one of the methods are plotted with empty circles/squares.}
\label{fig:BGX}
\end{figure}

Based on Figure \ref{fig:ElbowRule}, $\kappa_{1} = (1.5)\|\hat{\boldsymbol{\beta}}_{\text{L}}\|$ was chosen as our optimal constraint level. Note that at $\kappa_{1}$, the $\ell_{1}$ norm of the estimated coefficient vector is $1.5$ times that of the naive estimate, while selecting only half as many covariates. Figure \ref{fig:BGX} illustrates this by plotting the coefficient estimates of the naive lasso and the conditional scores lasso. For comparision, the covariates selected by both methods are plotted with filled circles/squares, whereas those selected by only one of the methods are plotted with empty circles/squares. The impact of the measurement error correction clearly is to amplify the coefficients of some seemingly important genes, while the naive lasso has many coefficients with magnitudes of the same order. This is an analogue to the measurement error attenuation in standard linear regression.

\par
\setcounter{chapter}{7}
\setcounter{equation}{0} 
\noindent {\bf 7. Discussion}

In this paper, we have shown how linear regression with the lasso is affected by additive measurement error. In particular, standard results for consistency of estimation and covariate selection do no longer hold when the covariates are subject to measurement error. A simple correction method was considered, studied earlier by Loh and Wainwright (2012). Our finite sample results show conditions under which this corrected 
lasso will be a sign consistent covariate selector. Asymptotically, sign consistent covariate selection with the corrected lasso requires conditions very similar to the lasso in the absence of measurement error. In contrast, asymptotically sign consistent covariate selection with the naive lasso essentially requires the relevant and the irrelevant covariates to be uncorrelated. We have also suggested a conditional scores approach for correcting for measurement error in $\ell_{1}$-constrained GLMs, which shows promising empirical results. Using the iteration scheme suggested by Loh and Wainwright (2012) for linear models, corrected lasso estimates for GLMs are computed efficiently even when $p>>n$.

The simulation results presented confirm that ignoring measurement error may yield a large number of false positive selections. The same is observed by Rosenbaum and Tsybakov (2010) for censored and missing data. Correction for measurement yields a much sparser fit, while finding almost as many of the relevant covariates. Our example application with microarray data agree well with the simulations. Measurement error correction yields a substantially sparser model, and the elbow rule (Figure \ref{fig:ElbowRule}) works very well for finding a good constraint level for the conditional scores lasso.


\noindent {\large\bf Acknowledgment}

The authors thank Po-Ling Loh and Martin J. Wainwright for sharing Matlab code used to compute the corrected lasso estimates. Thanks to Sylvia Richardson for discussions and for suggesting the use of BGX output data, and to Bin Yu for discussions. This work has been supported by funding from the Norwegian Research Council through the centre Statistics for Innovation in Oslo.
\par


\noindent{\large\bf References}
\begin{description}
\item
Ayers, K.L. and Cordell, H.J. (2010). SNP selection in genome-wide and candidate gene studies via penalized logistic regression. {\it Genet. Epidemiol.} {\bf 34}, 879-891.
\item
Benjamini, Y. and Speed, T. P. (2012). Summarizing and correcting the GC content bias in high-throughput sequencing. {\it Nucl. Acids Res.} {\bf 40}, e72.
\item
Bühlmann, P. and van de Geer, S. (2011). {\it Statistics for high-dimensional data.} Springer, Heidelberg.
\item
Candès, E. and Tao, T. (2007). The Dantzig selector: Statistical estimation when $p$ is much larger than $n$. {\it Ann. Statist.} {\bf 35}, 2313-2351.
\item
Carroll, R. J., Ruppert, D., Stefanski, L. A. and Crainiceanu, C. M. (2006). {\it Measurement error in nonlinear models.} Chapman and Hall/CRC, Boca Raton.
\item
Chen, Y. and Caramanis, C. (2013). Noisy and missing data regression: Distribution-oblivious support recovery. {\it J. Mach. Learn. Res. W\&CP.} {\bf 28}, 383-391.
\item
Duchi, J., Shalev-Shwartz, S., Singer, Y. and Chandra, T. (2008). Efficient projections onto the $\ell_{1}$-ball for learning in high dimensions. {\it Proceedings of the $25$th international conference on machine learning.} 272-279.
\item
Fan, J. and Li, R. (2001). Variable selection via nonconcave penalized likelihood and its oracle properties. {\it J. Amer. Statist. Assoc.} {\bf 96}, 1348-1360.
\item
Friedman, J., Hastie, T. and Tibshirani, R. (2010). Regularization Paths for
Generalized Linear Models via Coordinate Descent. {\it J. Stat. Softw.} {\bf 33}, 1-22.
\item
Hanfelt, J.J. and Liang, K.-Y. (1997). Approximate Likelihoods for Generalized Linear Errors-in-variables Models. {\it J. Roy. Statist. Soc. Ser. B.} {\bf 59}, 627-637.
\item
Hein, A.-M.K., Richardson, S., Causton, H.C., Ambler, G.K. and Green,
P.J. (2005). BGX: a fully Bayesian integrated approach to the analysis of
Affymetrix GeneChip data. {\it Biostatistics.} {\bf 6}, 349-373.
\item
Hoerl, A. E. and Kennard, R. W. (1970). Ridge regression: Biased estimation for nonorthogonal problems. {\it Technometrics.} {\bf 12}, 55-67.
\item
Huang, H.-C., Hsu, N.-J., Theobald, D.M. and Breidt, F.J. (2010). Spatial Lasso With Applications to GIS Model Selection. {\it J. Comput. Graph. Statist.} {\bf 19}, 963-983.
\item
Knight, K. and Fu, W. (2000). Asymptotics of lasso-type estimators. {\it Ann. Statist.} {\bf 28}, 1356-1378.
\item
Loh, P.-L. and Wainwright, M. J. (2012). High-dimensional regression with noisy and missing data: Provable guarantees with non-convexity. {\it Ann. Statist.} {\bf 40}, 1637-1664.
\item
Liang, H. and Li, R. (2009). Variable selection for partially linear models with measurement errors. {\it J. Amer. Statist. Assoc.} {\bf 104}, 234-248.
\item
Ma, Y. and Li, R. (2010). Variable selection in measurement error models. {\it Bernoulli.} {\bf 16}, 274-300.
\item
Meinshausen, N. and Bühlmann, P. (2010). Stability selection. {\it J. Roy. Statist. Soc. Ser. B.} {\bf 72}, 417-473.
\item
Purdom, E. and Holmes, S. P. (2005). Error distribution for gene expression data. {\it Stat. Appl. Genet. Mol. Biol.} {\bf 4}.
\item
Rocke, D. M. and Durbin, B. (2001). A model for measurement error for gene expression arrays. {\it J. Comput. Biol.} {\bf 8}, 557-569.
\item
Rosenbaum, M. and Tsybakov, A. B. (2010). Sparse recovery under matrix uncertainty. {\it Ann. Statist.} {\bf 38}, 2620-2651.
\item
Rosenbaum, M. and Tsybakov, A. B. (2013). Improved matrix uncertainty selector. {\it IMS Collections. From probability to statistics and back: High-dimensional models and processes.} {\bf 9}, 276-290.
\item
Stefanski, L. A. and Carroll, R. J. (1987). Conditional scores and optimal scores for generalized linear measurement-error models. {\it Biometrika.} {\bf 74}, 703-716.
\item
Tibshirani, R. (1996). Regression shrinkage and selection via the lasso. {\it J. Roy. Statist. Soc. Ser. B.} {\bf 58}, 267-288.
\item
Wu, T.T., Chen, Y.F., Hastie, T., Sobel, E., and Lange, K. (2009). Genome-wide association analysis by lasso penalized logistic regression. {\it Bioinformatics.} {\bf 25}, 714-721.
\item
Xu, Q. and You, J. (2007). Covariate selection for linear errors-in-variables regression models. {\it Comm. Statist. Theory Methods.} {\bf 36}, 375-386.
\item
Zhao, P. and Yu, B. (2006). On Model Selection Consistency of Lasso. {\it J. Mach. Learn. Res.} {\bf 7}, 2541-2563.
\item
Zou, H. (2006). The adaptive lasso and its oracle properties. {\it J. Amer. Statist. Assoc.} {\bf 101}, 1418-1429.
\end{description}

\vskip .65cm
\noindent
Øystein Sørensen\\
Department of Biostatistics, Institute of Basic Medical Sciences, University of Oslo,
Oslo, Norway\\
E-mail: oystein.sorensen@medisin.uio.no\\
Phone: +47 98806283
\vskip 2pt

\noindent
Arnoldo Frigessi\\
Department of Biostatistics, Institute of Basic Medical Sciences, University of Oslo,
Oslo, Norway\\
E-mail: arnoldo.frigessi@medisin.uio.no
\vskip 2pt

\noindent
Magne Thoresen\\
Department of Biostatistics, Institute of Basic Medical Sciences, University of Oslo,
Oslo, Norway\\
E-mail: magne.thoresen@medisin.uio.no
\vskip 2pt


\fontsize{10.95}{14pt plus.8pt minus .6pt}\selectfont
\vspace{0.8pc}
\centerline{\large\bf SUPPLEMENTARY MATERIAL}
\vspace{2pt}

\section{Regularity conditions}
We assume fixed true covariates which satisfy 
\begin{equation}\label{eq:Convergence1SigmaX}
(1/n)\mathbf{X}'\mathbf{X} = \mathbf{C}_{xx} \to \boldsymbol{\Sigma}_{xx}, \text{ as } n\to \infty
\end{equation}
and 
\begin{equation}\label{eq:Convergence2SigmaX}
(1/n) \underset{1\leq i \leq n}{\text{max}}\left( \mathbf{x}_{i}'\mathbf{x}_{i}\right) \to 0, \text{ as } n\to \infty,
\end{equation}
where $\boldsymbol{\Sigma}_{xx}$ is a positive definite matrix. 

The random measurement errors are assumed normally distributed with mean zero and covariance $\boldsymbol{\Sigma}_{uu}$. It follows (Anderson (2003, Th. 3.4.4)) that the limiting distribution of $n^{1/2}\left(\mathbf{C}_{uu} - \boldsymbol{\Sigma}_{uu} \right)$ is normal with mean $\mathbf{0}$ and covariances $\left(\boldsymbol{\Sigma}_{uu}\right)_{ik}\left(\boldsymbol{\Sigma}_{uu}\right)_{jl} + \left(\boldsymbol{\Sigma}_{uu}\right)_{il}\left(\boldsymbol{\Sigma}_{uu}\right)_{jk}$, where $\left(\boldsymbol{\Sigma}_{uu}\right)_{ik}$ is the $\left(i,k\right)$th element of $\boldsymbol{\Sigma}_{uu}$ and $i,j,k,l \in \left\{1,\dots,p\right\}$. Now the convergences
\begin{equation}\label{eq:Convergence1SigmaU}
\mathbf{C}_{uu} \to \boldsymbol{\Sigma}_{uu}, \text{ as } n \to \infty
\end{equation}
and
\begin{equation}\label{eq:Convergence2SigmaU}
(1/n) \underset{1\leq i \leq n}{\text{max}}\left( \mathbf{u}_{i}'\mathbf{u}_{i}\right) \to 0, \text{ as } n\to \infty,
\end{equation}
hold with probability $1$.

It follows from (\ref{eq:Convergence1SigmaX})-(\ref{eq:Convergence2SigmaU}) that the measurements $\mathbf{W} = \mathbf{X} + \mathbf{U}$ satisfy
\begin{equation}\label{eq:Convergence1SigmaW}
\mathbf{C}_{ww} \to \boldsymbol{\Sigma}_{ww}, \text{ as } n \to \infty
\end{equation}
and
\begin{equation}\label{eq:Convergence2SigmaW}
(1/n) \underset{1\leq i \leq n}{\text{max}}\left( \mathbf{w}_{i}'\mathbf{w}_{i}\right) \to 0, \text{ as } n\to \infty,
\end{equation}
with probability $1$. Also, the limiting distribution of $n^{1/2}\left(\mathbf{C}_{ww} - \boldsymbol{\Sigma}_{ww} \right)$ will have mean zero and finite covariances.

Regularity conditions like these have also been assumed by, e.g., Knight and Fu (2000) and Zhao and Yu (2006).
\par

\section{Karush-Kuhn-Tucker Conditions}
\setcounter{equation}{0}

We introduce the new coefficient $\boldsymbol{\gamma} = \boldsymbol{\beta} - \boldsymbol{\beta}^{0}$, which yields the naive lasso on the form
\begin{equation}
\hat{\boldsymbol{\gamma}} = \text{arg}~ \underset{\boldsymbol{\gamma}}{\text{min}} \left(- \frac{2}{n}\boldsymbol{\epsilon}'\mathbf{W} \boldsymbol{\gamma} + \boldsymbol{\gamma}' \mathbf{C}_{ww}\boldsymbol{\gamma} + 2 \boldsymbol{\gamma}' \mathbf{C}_{wu} \boldsymbol{\beta}^{0} + \lambda \left\| \boldsymbol{\gamma} + \boldsymbol{\beta}^{0}\right\|_{1} \right),
\label{eq:ReparNaiveLasso}
\end{equation}
where we have removed all terms which are constant in $\boldsymbol{\gamma}$. Taking derivatives, we arrive at the Karush-Kuhn-Tucker conditions for the naive Lasso.

\noindent{\bf Lemma 1.} \textit{$\hat{\boldsymbol{\gamma}} = \hat{\boldsymbol{\beta}} - \boldsymbol{\beta}^{0}$ is a solution to (\ref{eq:ReparNaiveLasso}) if and only if
\begin{equation*}
- \frac{2}{n}\boldsymbol{\epsilon}' \mathbf{W} + 2 \mathbf{C}_{ww}  \hat{\boldsymbol{\gamma}} + 2  \mathbf{C}_{wu}  = - \lambda \hat{\boldsymbol{\tau}},
\end{equation*}
where $\hat{\boldsymbol{\tau}} \in \mathbb{R}^{p}$ satisfies $\left\|\hat{\boldsymbol{\tau}}\right\|_{\infty} \leq 1$ and $\hat{\tau}_{j} = \text{sign}\left( \hat{\beta}_{j} \right)$ for $j$ such that $\hat{\beta}_{j} \neq 0$.
}

The same change of variables for the corrected lasso yields
\begin{align}\label{eq:ReparCorrLasso}
\hat{\boldsymbol{\gamma}} =  \underset{\boldsymbol{\gamma}~:~ \left\|\boldsymbol{\gamma} + \boldsymbol{\beta}^{0}\right\|_{1} \leq R}{\text{arg~min}} \bigg\{&- \frac{2}{n}\boldsymbol{\epsilon}'\mathbf{W} \boldsymbol{\gamma} + \boldsymbol{\gamma}' \left(\mathbf{C}_{ww} - \boldsymbol{\Sigma}_{uu} \right)\boldsymbol{\gamma} \\ \nonumber
&+ 2 \boldsymbol{\gamma}' \left(\mathbf{C}_{wu}  - \boldsymbol{\Sigma}_{uu} \right) \boldsymbol{\beta}^{0} + \lambda \left\| \boldsymbol{\gamma} + \boldsymbol{\beta}^{0}\right\|_{1} \bigg\}.
\end{align}
Due to the additional constraint $\left\|\boldsymbol{\gamma} + \boldsymbol{\beta}^{0}\right\|_{1} \leq R$ added because of non-convexity, the KKT conditions can only characterize critical points in the interior of this domain. A critical point on the boundary may not have a zero subgradient. However, under the assumptions of Loh and Wainwright (2012), for sufficiently large $n$, all local optima lie in a small $\ell_{1}$-ball around $\boldsymbol{\beta}^{0}$. We assume that $R$ is chosen large enough such that $\left\|\boldsymbol{\gamma} + \boldsymbol{\beta}^{0}\right\|_{1} < R$ for all these optima, while $R$ is small enough to avoid the trivial solutions for which one or more component of $\hat{\boldsymbol{\gamma}}$ is $\pm \infty$.

\noindent{\bf Lemma 2.} \textit{Assume $\hat{\boldsymbol{\gamma}} = \hat{\boldsymbol{\beta}} - \boldsymbol{\beta}^{0}$ is a critical point of (\ref{eq:ReparCorrLasso}). If $\hat{\boldsymbol{\gamma}}$ lies in the interior of the feasible set, i.e., $\left\|\hat{\boldsymbol{\gamma}} + \boldsymbol{\beta}^{0}\right\|_{1} < R$, then
\begin{equation*}
- \frac{2}{n}\boldsymbol{\epsilon}' \mathbf{W} + 2 \left(\mathbf{C}_{ww} - \boldsymbol{\Sigma}_{uu} \right) \hat{\boldsymbol{\gamma}} + 2 \left( \mathbf{C}_{wu}  - \boldsymbol{\Sigma}_{uu} \right) = - \lambda \hat{\boldsymbol{\tau}},
\end{equation*}
where $\hat{\boldsymbol{\tau}} \in \mathbb{R}^{p}$ is as defined in Lemma 1.
}
\par

\section{Proof of Proposition 1}
\setcounter{equation}{0}
The following basic inequality for the lasso follows by definition (Bühlmann and van de Geer (2011)):
\begin{equation*}
\left(1/n\right) \left\|\mathbf{y} - \mathbf{W} \hat{\boldsymbol{\beta}}\right\|_{2}^{2} + \lambda \left\|\hat{\boldsymbol{\beta}}\right\|_{1} \leq \left(1/n\right) \left\|\mathbf{y} - \mathbf{W} \boldsymbol{\beta}^{0}\right\|_{2}^{2} + \lambda \left\|\boldsymbol{\beta}^{0}\right\|_{1}.
\end{equation*}
Reorganizings terms, we arrive at
\begin{equation}\label{eq:BasicIneq}
(1/n)\left\|\mathbf{W} \left( \hat{\boldsymbol{\beta}} - \boldsymbol{\beta}^{0}\right) \right\|_{2}^{2} + \lambda \left\|\hat{\boldsymbol{\beta}}\right\|_{1} \leq (2/n) \left( \boldsymbol{\epsilon} - \mathbf{U} \boldsymbol{\beta}^{0}\right)'\mathbf{W} \left(\hat{\boldsymbol{\beta}} - \boldsymbol{\beta}^{0}\right) + \lambda \left\|\boldsymbol{\beta}^{0}\right\|_{1}.
\end{equation}
Under the noise bound (3.1), it is clear that
\begin{equation*}
(2/n) \left( \boldsymbol{\epsilon} - \mathbf{U} \boldsymbol{\beta}^{0}\right)'\mathbf{W} \left(\hat{\boldsymbol{\beta}} - \boldsymbol{\beta}^{0}\right) \leq (2/n) \left\|\left(\boldsymbol{\epsilon} - \mathbf{U} \boldsymbol{\beta}^{0}\right) \mathbf{W}\right\|_{\infty} \left\|\hat{\boldsymbol{\beta}} - \boldsymbol{\beta}^{0}\right\|_{1} \leq \lambda_{0} \left\|\hat{\boldsymbol{\beta}} - \boldsymbol{\beta}^{0}\right\|_{1},
\end{equation*}
which, inserted into (\ref{eq:BasicIneq}) yields
\begin{equation*}
(1/n)\left\|\mathbf{W} \left( \hat{\boldsymbol{\beta}} - \boldsymbol{\beta}^{0}\right) \right\|_{2}^{2} + \lambda \left\|\hat{\boldsymbol{\beta}}\right\|_{1} \leq \lambda_{0} \left\|\hat{\boldsymbol{\beta}} - \boldsymbol{\beta}^{0}\right\|_{1} + \lambda\left\|\boldsymbol{\beta}^{0}\right\|_{1}.
\end{equation*}
Now use the inequality
\begin{equation*}
\left\|\hat{\boldsymbol{\beta}}\right\|_{1} \geq \left\|\boldsymbol{\beta}_{S_{0}}^{0}\right\|_{1} - \left\|\hat{\boldsymbol{\beta}}_{S_{0}} -\boldsymbol{\beta}^{0}_{S_{0}} \right\|_{1} + \left\|\hat{\boldsymbol{\beta}}_{S_{0}^{c}} \right\|_{1},
\end{equation*}
the equality
\begin{equation}
\left\|\hat{\boldsymbol{\beta}} - \boldsymbol{\beta}^{0}\right\|_{1} = \left\|\hat{\boldsymbol{\beta}}_{S_{0}} - \boldsymbol{\beta}^{0}_{S_{0}}\right\|_{1} + \left\|\hat{\boldsymbol{\beta}}_{S_{0}^{c}}\right\|_{1}
\label{eq:IntStepEquality}
\end{equation}
and $\lambda \geq 2\lambda_{0}$, to obtain
\begin{equation}
(2/n) \left\| \mathbf{W} \left( \hat{\boldsymbol{\beta}} - \boldsymbol{\beta}^{0}\right)\right\|_{2}^{2} + \lambda \left\|\hat{\boldsymbol{\beta}}_{S_{0}^{c}} \right\|_{1} \leq 3 \lambda \left\|\hat{\boldsymbol{\beta}}_{S_{0}} - \boldsymbol{\beta}^{0}_{S_{0}}\right\|_{1}.
\label{eq:IntStep1}
\end{equation}

Inequality (\ref{eq:IntStep1}) shows that $\|\hat{\boldsymbol{\beta}}_{S_{0}^{c}} \|_{1} \leq 3\|\hat{\boldsymbol{\beta}}_{S_{0}} - \boldsymbol{\beta}^{0}_{S_{0}}\|_{1}$. That is, the vector $\hat{\boldsymbol{\beta}} - \boldsymbol{\beta}^{0}$ is among the vectors to which the compatibility condition applies, for the index set $S_{0}$. Next, use (\ref{eq:IntStepEquality}) again in (\ref{eq:IntStep1}) to obtain
\begin{equation}
(2/n) \left\| \mathbf{W} \left( \hat{\boldsymbol{\beta}} - \boldsymbol{\beta}^{0}\right)\right\|_{2}^{2} +  \lambda \left\|\hat{\boldsymbol{\beta}} - \boldsymbol{\beta}^{0}\right\|_{1} \leq 4 \lambda \left\|\hat{\boldsymbol{\beta}}_{S_{0}} - \boldsymbol{\beta}^{0}_{S_{0}}\right\|_{1}.
\label{eq:IntStep2}
\end{equation}

Since $\hat{\boldsymbol{\beta}} - \boldsymbol{\beta}^{0}$ is in the set of vectors for which the compatibility condition holds on $S_{0}$, we have
\begin{equation*}
\left\|\hat{\boldsymbol{\beta}}_{S_{0}} - \boldsymbol{\beta}^{0}_{S_{0}}\right\|_{1} \leq s_{0}^{1/2}\phi_{0}^{-1} n^{-1/2}\left\|\mathbf{W}\left( \hat{\boldsymbol{\beta}} - \boldsymbol{\beta}^{0}\right)\right\|_{2} .
\end{equation*}
Using this and the inequality $4uv \leq 4u^{2} + v^{2}$ in (\ref{eq:IntStep2}), we arrive at
\begin{equation*}
\left\|\mathbf{W} \left( \hat{\boldsymbol{\beta}} - \boldsymbol{\beta}^{0}\right) \right\|_{2}^{2} + \lambda \left\| \hat{\boldsymbol{\beta}} - \boldsymbol{\beta}_{S_{0}}^{0}\right\|_{1} \leq 4 \lambda^{2} s_{0} /\phi_{0}^{2}.
\label{eq:Consistency}
\end{equation*}

\par

\section{Proof of Proposition 2}

This proof goes along the lines of the proof of Theorem 1 in Knight and Fu (2000), but with the addition of measurement error. We start with the naive Lasso after reparametrization, and denote its Lagrange function by
\begin{equation}\label{eq:AsymptoticsLagrange}
\mathcal{L}_{n} (\boldsymbol{\gamma}) = -\frac{2}{\sqrt{n}} \boldsymbol{\gamma}' \frac{\mathbf{W}}{\sqrt{n}} \boldsymbol{\epsilon} + \boldsymbol{\gamma}' \mathbf{C}_{ww} \boldsymbol{\gamma} +2 \boldsymbol{\gamma}'\mathbf{C}_{wu}\boldsymbol{\beta}^{0} + \lambda\|\boldsymbol{\gamma} + \boldsymbol{\beta}^{0} \|_{1}.
\end{equation}
Note that 
\begin{equation*}
\frac{2}{\sqrt{n}} \frac{\mathbf{W}}{\sqrt{n}} \boldsymbol{\epsilon} \overset{d}{\to} \mathcal{N}(\mathbf{0}, (4/n)\sigma^{2} \boldsymbol{\Sigma}_{ww}).
\end{equation*}
That is, the first term in (\ref{eq:AsymptoticsLagrange}) converges in distribution to a normally distributed quantity whose variance goes to zero as $1/n$, which is equivalent to convergence in probability to zero. Combining this result with the assumption that $\lambda \to 0$ as $n\to \infty$, yields
\begin{equation*}
\mathcal{L}_{n} (\boldsymbol{\gamma}) \overset{p}{\to} \mathcal{L} (\boldsymbol{\gamma}) = \boldsymbol{\gamma}' \boldsymbol{\Sigma}_{ww} \boldsymbol{\gamma} +2 \boldsymbol{\gamma}'\boldsymbol{\Sigma}_{uu}\boldsymbol{\beta}^{0}.
\end{equation*}
Since $\mathcal{L}_{n}(\boldsymbol{\gamma})$ is convex, it follows that (Knight and Fu (2000))
\begin{equation*}
\text{arg}\underset{\boldsymbol{\gamma}}{\text{min}}\{\mathcal{L}_{n} (\boldsymbol{\gamma})\} \overset{p}{\to} \text{arg}\underset{\boldsymbol{\gamma}}{\text{min}}\{\mathcal{L} (\boldsymbol{\gamma})\}.
\end{equation*}
The minimum of $\mathcal{L}(\boldsymbol{\gamma})$ is easily found, and accordingly,
\begin{equation*}
\hat{\boldsymbol{\gamma}} \overset{p}{\to} - \boldsymbol{\Sigma}_{ww}^{-1} \boldsymbol{\Sigma}_{uu} \boldsymbol{\beta}^{0}.
\end{equation*}
The result follows immediately.

\par

\section{Proof of Theorem 1}
\setcounter{equation}{0}
We follow the structure of the proof by Zhao and Yu (2006), who proved the corresponding result in the absence of measurement error. Consider the naive lasso, and note that (Zhao and Yu (2006))
\begin{equation*}
\left\{\text{sign}\left(\boldsymbol{\beta}_{S_{0}}^{0} \right) \hat{\boldsymbol{\gamma}}_{S_{0}} > - \left|\boldsymbol{\beta}_{S_{0}}^{0} \right| \right\} ~ \Rightarrow ~ \left\{\text{sign}\left(\hat{\boldsymbol{\beta}}_{S_{0}} \right) = \text{sign}\left(\boldsymbol{\beta}_{S_{0}}^{0} \right) \right\}
\end{equation*}
and
\begin{equation*}
\hat{\boldsymbol{\gamma}}_{S_{0}^{c}} = \boldsymbol{0}  ~\Rightarrow ~ \hat{\boldsymbol{\beta}}_{S_{0}^{c}} = \boldsymbol{0}.
\end{equation*}
Thus, by the Karush-Kuhn-Tucker conditions for the naive lasso (Lemma 1), if a solution $\hat{\boldsymbol{\gamma}}$ exists, and the following three conditions hold:
\begin{align}\label{eq:KKTNaive1}
- \frac{\mathbf{W}_{S_{0}}'}{\sqrt{n}} \boldsymbol{\epsilon} + \sqrt{n} \mathbf{C}_{ww}\left(S_{0},S_{0} \right) \hat{\boldsymbol{\gamma}}_{S_{0}} + \sqrt{n} C_{wu}\left(S_{0},S_{0} \right) \boldsymbol{\beta}_{S_{0}}^{0} = - \frac{\lambda\sqrt{n}}{2 } \text{sign}\left(\boldsymbol{\beta}^{0}_{S_{0}}\right) \\ \label{eq:KKTNaive2}
\left|\hat{\boldsymbol{\gamma}}_{S_{0}} \right| < \left|\boldsymbol{\beta}_{S_{0}}^{0} \right| \\ \label{eq:KKTNaive3}
\left|- \frac{W_{S_{0}^{c}}'}{\sqrt{n}} \boldsymbol{\epsilon} + \sqrt{n} \mathbf{C}_{ww}\left( S_{0}^{c},S_{0}\right) \hat{\boldsymbol{\gamma}}_{S_{0}} + \sqrt{n} \mathbf{C}_{wu}\left(S_{0}^{c},S_{0} \right) \boldsymbol{\beta}_{S_{0}}^{0} \right| \leq \frac{\lambda \sqrt{n}}{2} \boldsymbol{1},
\end{align}
then $\text{sign}(\hat{\boldsymbol{\beta}}_{S_{0}} ) = \text{sign}(\boldsymbol{\beta}_{S_{0}}^{0} )$ and $\text{sign}(\hat{\boldsymbol{\beta}}_{S_{0}^{c}} ) = \mathbf{0}$.

Event $A$ implies the existence of $|\hat{\boldsymbol{\gamma}}_{S_{0}}| < |\boldsymbol{\beta}_{S_{0}}^{0}|$ such that
\begin{equation*}
\left|\mathbf{Z}_{1}' \boldsymbol{\epsilon} - \mathbf{Z}_{2} \boldsymbol{\beta}_{S_{0}}^{0} \right| = \sqrt{n}\left(\left|\hat{\boldsymbol{\gamma}}_{S_{0}} \right| - \frac{\lambda}{2} \left|\mathbf{C}_{ww}\left(S_{0},S_{0} \right)^{-1} \text{sign}\left(\boldsymbol{\beta}_{S_{0}}^{0} \right) \right| \right).
\end{equation*}
Buth then there must also exist $|\hat{\boldsymbol{\gamma}}_{S_{0}}| < |\boldsymbol{\beta}_{S_{0}}^{0}|$ such that
\begin{equation*}
\mathbf{Z}_{1}' \boldsymbol{\epsilon} - \mathbf{Z}_{2} \boldsymbol{\beta}_{S_{0}}^{0} = \sqrt{n} \left( \hat{\boldsymbol{\gamma}}_{S_{0}}  - \frac{\lambda}{2} \mathbf{C}_{ww}\left(S_{0},S_{0} \right)^{-1} \text{sign}\left(\boldsymbol{\beta}_{S_{0}}^{0} \right) \right),
\end{equation*} 
which essentially means choosing the appropriate signs of the elements of $\hat{\boldsymbol{\gamma}}_{S_{0}}$. Multiplying through by $\mathbf{C}_{ww}(S_{0},S_{0})$ and reorganizing terms, we get (\ref{eq:KKTNaive1}). Thus, $A$ ensures that (\ref{eq:KKTNaive1}) and (\ref{eq:KKTNaive2}) are satisfied. Next, adding and subtracting $\sqrt{n}\mathbf{C}_{ww}(S_{0}^{c},S_{0})\hat{\boldsymbol{\gamma}}_{S_{0}}$ to the left-hand side of event $B$ and then using the triangle inequaltity, yields
\begin{align*}
\left|- \frac{\mathbf{W}_{S_{0}^{c}}}{\sqrt{n}} \boldsymbol{\epsilon} + \sqrt{n} + \mathbf{C}_{ww}(S_{0}^{c},S_{0}) \hat{\boldsymbol{\gamma}}_{S_{0}}+\sqrt{n} \mathbf{C}_{wu}(S_{0}^{c},S_{0}) \boldsymbol{\beta}_{S_{0}}^{0} \right|- \\
\bigg|- \mathbf{C}_{ww}(S_{0}^{c},S_{0}) \mathbf{C}_{ww}(S_{0},S_{0})^{-1} \frac{\mathbf{W}_{S_{0}}'}{\sqrt{n}} \boldsymbol{\epsilon} + \sqrt{n} \mathbf{C}_{ww}(S_{0}^{c},S_{0}) \mathbf{C}_{ww}(S_{0},S_{0})^{-1} \mathbf{C}_{wu}(S_{0},S_{0}) \boldsymbol{\beta}_{S_{0}}^{0}\\
+ \sqrt{n} \mathbf{C}_{ww}(S_{0}^{c},S_{0}) \hat{\boldsymbol{\gamma}}_{S_{0}} \bigg| \leq \frac{\lambda \sqrt{n}}{2}\left(1-\theta\right) \boldsymbol{1}.
\end{align*}
The second term on the left-hand side of this expression is the left-hand side of (\ref{eq:KKTNaive1}) multiplied by $\mathbf{C}_{ww}(S_{0}^{c},S_{0}) \mathbf{C}_{ww}(S_{0},S_{0})^{-1}$. It can thus be replaced by the right-hand side of (\ref{eq:KKTNaive1}) multiplied by this factor. This yields
\begin{align*}
\left|- \frac{\mathbf{W}_{S_{0}^{c}}}{\sqrt{n}} \boldsymbol{\epsilon} + \sqrt{n} + \mathbf{C}_{ww}(S_{0}^{c},S_{0}) \hat{\boldsymbol{\gamma}}_{S_{0}}+\sqrt{n} \mathbf{C}_{wu}(S_{0}^{c},S_{0}) \boldsymbol{\beta}_{S_{0}}^{0} \right|- \\
\left|\frac{\lambda\sqrt{n}}{2 } \mathbf{C}_{ww}(S_{0}^{c},S_{0}) \mathbf{C}_{ww}(S_{0},S_{0})^{-1} \text{sign}\left(\boldsymbol{\beta}^{0}_{S_{0}}\right)\right| \leq \frac{\lambda \sqrt{n}}{2}\left(1-\theta\right) \boldsymbol{1},
\end{align*}
which implies, due to the IC-ME,
\begin{align*}
\left|- \frac{\mathbf{W}_{S_{0}^{c}}}{\sqrt{n}} \boldsymbol{\epsilon} + \sqrt{n}  \mathbf{C}_{ww}(S_{0}^{c},S_{0}) \hat{\boldsymbol{\gamma}}_{S_{0}}+\sqrt{n} \mathbf{C}_{wu}(S_{0}^{c},S_{0}) \boldsymbol{\beta}_{S_{0}}^{0} \right|
 \leq \frac{\lambda \sqrt{n}}{2} \boldsymbol{1}.
\end{align*}
This is indeed (\ref{eq:KKTNaive3}). Altogether, $A$ implies (\ref{eq:KKTNaive1}) and (\ref{eq:KKTNaive2}), while $B|A$ implies (\ref{eq:KKTNaive3}).

For the asymptotic result, define the vectors
\begin{align*}
&\mathbf{z} = \mathbf{C}_{ww}(S_{0},S_{0})^{-1}  \frac{\mathbf{W}_{S_{0}}'}{\sqrt{n}}\boldsymbol{\epsilon} \\
&\mathbf{a} = \left| \boldsymbol{\beta}_{S_{0}}^{0}\right| - \left|\mathbf{C}_{ww}(S_{0},S_{0})^{-1} \mathbf{C}_{wu}(S_{0},S_{0}) \boldsymbol{\beta}_{S_{0}}^{0} \right| \\
&\mathbf{b} = \mathbf{C}_{ww}(S_{0},S_{0})^{-1} \text{sign}\left(\boldsymbol{\beta}_{S_{0}}^{0} \right) \\
&\boldsymbol{\zeta} = \left(\mathbf{C}_{ww}(S_{0}^{c},S_{0}) \mathbf{C}_{ww}(S_{0},S_{0})^{-1} \frac{\mathbf{W}_{S_{0}}'}{\sqrt{n}} - \frac{\mathbf{W}_{S_{0}^{c}}'}{\sqrt{n}} \right)\boldsymbol{\epsilon} \\
&\mathbf{f} = \left( \mathbf{C}_{ww}(S_{0}^{c},S_{0}) \mathbf{C}_{ww}(S_{0},S_{0})^{-1} \mathbf{C}_{wu}(S_{0},S_{0}) - \mathbf{C}_{wu}(S_{0}^{c},S_{0})\right)\boldsymbol{\beta}_{S_{0}}^{0}.
\end{align*}
We have
\begin{align*}
&1- P(A\cap B) \leq P(A^{c}) +P(B^{c}) 
\leq \\
&\qquad \sum_{j=1}^{s_{0}} P\left(\left|z_{j} \right| \geq \sqrt{n}\left(a_{j} - \frac{\lambda}{2} b_{j} \right)\right) + \sum_{j=1}^{p-s_{0}}P\left(\left| \zeta_{j} - \sqrt{n}f_{j}\right| \geq \frac{\lambda \sqrt{n}}{2} \left(1-\theta\right) \right).
\end{align*}
It is clear that
\begin{align*}
\mathbf{z} \overset{d}{\to} \mathcal{N}\left(\mathbf{0}, \sigma^{2} \mathbf{C}_{ww}(S_{0},S_{0})^{-1} \right), \text{ as } n\to \infty.
\end{align*}
Hence, there exists a finite constant $k$ such that $E(z_{j})^{2} < k^{2}$ for $j=1,\dots,s_{0}$. Next, we have by assumption
\begin{align*}
\mathbf{a} \to \left| \boldsymbol{\beta}_{S_{0}}^{0}\right| - \left|\boldsymbol{\Sigma}_{ww}(S_{0},S_{0})^{-1} \boldsymbol{\Sigma}_{uu}(S_{0},S_{0}) \boldsymbol{\beta}_{S_{0}}^{0} \right|, \text{ as } n\to \infty,
\end{align*}
and 
\begin{align*}
\mathbf{b} \to \boldsymbol{\Sigma}_{ww}(S_{0},S_{0})^{-1} \text{sign}\left(\boldsymbol{\beta}_{S_{0}}^{0}\right), \text{ as } n \to \infty.
\end{align*}
Now using the assumption $\lambda = o(1)$, we get
\begin{align*}
P\left(A^{c}\right) &\leq \sum_{j=1}^{s_{0}} \left(1 - P\left(\frac{\left|z_{j}\right|}{k} < \frac{\sqrt{n}}{2k}a_{j}\left(1+o(1)\right) \right) \right)\\
&\leq \left(1 + o(1) \right)\sum_{j=1}^{s_{0}} \left(1 - \Phi\left(\frac{\sqrt{n}}{2s}a_{j} \left(1+o(1)\right) \right) \right) \\
&= o\left(\exp(-n^{c})\right),
\end{align*}
where we used the bound for the Gaussian tail probability
\begin{equation}
1 - \Phi(t) < t^{-1} \exp\left(-(1/2)t^{2} \right).
\label{eq:GaussianTail}
\end{equation}
Next, we note that
\begin{align*}
\zeta \overset{d}{\to} \mathcal{N}\left(\mathbf{0}, \sigma^{2}\left(\boldsymbol{\Sigma}_{ww}(S_{0}^{c},S_{0}^{c}) - \boldsymbol{\Sigma}_{ww}(S_{0}^{c},S_{0}) \boldsymbol{\Sigma}_{ww}(S_{0},S_{0})^{-1} \boldsymbol{\Sigma}_{ww}(S_{0},S_{0}^{c})\right) \right), \text{ as } n\to \infty.
\end{align*}
Next, we consider $\mathbf{f}$, and note that the limiting distribution of $\sqrt{n}\mathbf{C}_{wu} = \sqrt{n}( \mathbf{C}_{uu} + \mathbf{C}_{xu})$ as $n \to \infty$ is normal with mean $\sqrt{n}\boldsymbol{\Sigma}_{wu} = \sqrt{n}\boldsymbol{\Sigma}_{uu}$ and finite variances (Anderson (2003, Th. 3.4.4)). In addition, $\mathbf{C}_{ww} \to \boldsymbol{\Sigma}_{ww}$ as $n\to \infty$. Thus, applying Slutsky's theorem to the product of the matrices, the limiting distribution of 
\begin{align*}
\sqrt{n}\left(\mathbf{C}_{ww}(S_{0}^{c},S_{0}) \mathbf{C}_{ww}(S_{0},S_{0})^{-1} \mathbf{C}_{wu}(S_{0},S_{0}) - \mathbf{C}_{wu}(S_{0}^{c},S_{0}) \right)
\end{align*}
as $n\to \infty$ has mean 
\begin{align*}
\sqrt{n}\left(\boldsymbol{\Sigma}_{ww}(S_{0}^{c},S_{0}) \boldsymbol{\Sigma}_{ww}(S_{0},S_{0})^{-1} \boldsymbol{\Sigma}_{wu}(S_{0},S_{0}) - \boldsymbol{\Sigma}_{wu}(S_{0}^{c},S_{0}) \right) = \mathbf{0}
\end{align*}
and finite variances. The latter term equals zero by the MEC. Now
\begin{align*}
\sqrt{n} \mathbf{f} = \sqrt{n}\left( \mathbf{C}_{ww}(S_{0}^{c},S_{0}) \mathbf{C}_{ww}(S_{0},S_{0})^{-1} \mathbf{C}_{wu}(S_{0},S_{0}) - \mathbf{C}_{wu}(S_{0}^{c},S_{0})\right)\boldsymbol{\beta}_{S_{0}}^{0}
\end{align*}
is a vector in $\mathbb{R}^{p-s_{0}}$ whose elements are linear combinations of variables whose limiting distribution as $n\to \infty$ is normal with mean zero and finite variances. Accordingly, the limiting distribution of $\sqrt{n}\mathbf{f}$ as $n\to \infty$ is normal with mean zero and finite variances.

So again there exists a finite constant $k$ such that $E(\zeta_{j} - \sqrt{n}f_{j})^{2} < k^{2}$ for $j=1,\dots,(p-s_{0})$. Thus, when $\lambda n^{(1-c)/2} \to \infty$ for $c \in [0,1)$, we have
\begin{align*}
P(B^{c}) &\leq \sum_{j=1}^{p-s_{0}} \left(1 - P\left(\frac{\left|\zeta_{j} - \sqrt{n}f_{j} \right|}{k} < \frac{1}{k} \frac{\lambda \sqrt{n}}{2}\left( 1 - \theta\right) \right) \right) \\
&\leq \left( 1 + o(1)\right) \sum_{j=1}^{p-s_{0}} \left(1 - \Phi\left(\frac{1}{k} \frac{\lambda \sqrt{n}}{2} \left( 1 - \theta\right) \right) \right)\\
& = o\left( \exp(-n^{c})\right).
\end{align*}
It follows that $P(A \cap B) = 1 - o(\exp(-n^{c}))$.

\par

\section{Proof of Proposition 3}
\setcounter{equation}{0}

We consider now the Lasso with $\boldsymbol{\epsilon}=\boldsymbol{0}$. In this case, $\mathbf{y} = \mathbf{X}\boldsymbol{\beta}^{0}$, and the Lasso becomes
\begin{equation*}
\hat{\boldsymbol{\beta}} = \text{arg}\underset{\boldsymbol{\beta}}{\text{min}}\left\{\|\mathbf{W}\boldsymbol{\beta} - \mathbf{X}\boldsymbol{\beta}^{0} \|_{2}^{2} + \lambda \|\boldsymbol{\beta} \|_{1} \right\}.
\end{equation*}
We follow the proof of Bühlmann and van de Geer (2011, Th. 7.1), but also take measurement error into account.

\subsubsection*{Part 1}
The KKT conditions take the form
\begin{align}\label{eq:NoiselessKKT1}
&2 \mathbf{C}_{ww}(S_{0},S_{0}) (\hat{\boldsymbol{\beta}}_{S_{0}}-\boldsymbol{\beta}_{S_{0}}^{0}) + 2 \mathbf{C}_{ww}(S_{0},S_{0}^{c})\hat{\boldsymbol{\beta}}_{S_{0}^{c}} + 2 \mathbf{C}_{wu}(S_{0},S_{0}) \boldsymbol{\beta}_{S_{0}}^{0} = -\lambda\hat{\boldsymbol{\tau}}_{S_{0}} \\ \label{eq:NoiselessKKT2}
&2 \mathbf{C}_{ww}(S_{0}^{c},S_{0}) (\hat{\boldsymbol{\beta}}_{S_{0}}-\boldsymbol{\beta}_{S_{0}}^{0}) + 2 \mathbf{C}_{ww}(S_{0}^{c},S_{0}^{c}) \hat{\boldsymbol{\beta}}_{S_{0}^{c}} + 2 \mathbf{C}_{wu}(S_{0}^{c},S_{0}) \boldsymbol{\beta}_{S_{0}}^{0} = -\lambda\hat{\boldsymbol{\tau}}_{S_{0}^{c}},
\end{align}
where $\hat{\boldsymbol{\tau}} = (\hat{\boldsymbol{\tau}}_{S_{0}}', \hat{\boldsymbol{\tau}}_{S_{0}^{c}}')'$ has the property that $\|\hat{\boldsymbol{\tau}}\|_{\infty}\leq 1$ and $\hat{\tau}_{j} = \text{sign}(\hat{\beta}_{j})$ if $\beta_{j}\neq 0$. We multiply (\ref{eq:NoiselessKKT1}) by $\hat{\boldsymbol{\beta}}_{S_{0}^{c}}' \mathbf{C}_{ww}(S_{0}^{c},S_{0}) \mathbf{C}_{ww}(S_{0},S_{0})^{-1}$ and (\ref{eq:NoiselessKKT2}) by $\hat{\boldsymbol{\beta}}_{S_{0}^{c}}'$, and then subtract the first from the second, to get
\begin{align}\nonumber
&2 \hat{\boldsymbol{\beta}}_{S_{0}^{c}}' \left(\mathbf{C}_{ww}(S_{0}^{c},S_{0}^{c}) -  \mathbf{C}_{ww}(S_{0}^{c},S_{0}) \mathbf{C}_{ww}(S_{0},S_{0})^{-1} \mathbf{C}_{ww}(S_{0},S_{0}^{c})\right) \hat{\boldsymbol{\beta}}_{S_{0}^{c}} +\\ \nonumber
&\qquad 2 \hat{\boldsymbol{\beta}}_{S_{0}^{c}}' \left( \mathbf{C}_{wu}(S_{0}^{c},S_{0}) - \mathbf{C}_{ww}(S_{0}^{c},S_{0}) \mathbf{C}_{ww}(S_{0},S_{0})^{-1} \mathbf{C}_{wu}(S_{0},S_{0})\right)\boldsymbol{\beta}_{S_{0}}^{0} \\ \label{eq:NoiselessCondition}
&\qquad= \lambda\left( \hat{\boldsymbol{\beta}}_{S_{0}^{c}}' \mathbf{C}_{ww}(S_{0}^{c},S_{0}) \mathbf{C}_{ww}(S_{0},S_{0})^{-1} \hat{\boldsymbol{\tau}}_{S_{0}} - \hat{\boldsymbol{\beta}}_{S_{0}^{c}}' \hat{\boldsymbol{\tau}}_{S_{0}^{c}} \right)
\end{align}
The matrix term within the parantheses in the leftmost term is positive semidefinite, since it is the Schur complement of the positive semidefinite matrix $\mathbf{C}_{ww}$, in which the part $\mathbf{C}_{ww}(S_{0},S_{0})$ is positive definite, since $l_{0} < n$. Next, the term within the parantheses on the right-hand side is
\begin{align*}
&\hat{\boldsymbol{\beta}}_{S_{0}}' \mathbf{C}_{ww}(S_{0}^{c},S_{0}) \mathbf{C}_{ww}(S_{0},S_{0})^{-1} \hat{\boldsymbol{\tau}}_{S_{0}} - \|\hat{\boldsymbol{\beta}}_{S_{0}} \|_{1} \leq \\
& \left(\left\| \mathbf{C}_{ww}(S_{0}^{c},S_{0}) \mathbf{C}_{ww}(S_{0},S_{0})^{-1} \hat{\boldsymbol{\tau}}_{S_{0}}\right\|_{\infty} - 1 \right)\left\|\hat{\boldsymbol{\beta}}_{S_{0}^{c}}\right\|_{1} \leq 0.
\end{align*}
The last inequality follows from the IC-ME, and is strict whenever $\|\hat{\boldsymbol{\beta}}_{S_{0}^{c}}\|_{1}\neq 0$. Finally, the second term on the left-hand side of  (\ref{eq:NoiselessCondition}) is zero by assumption. Thus, if $\|\hat{\boldsymbol{\beta}}_{S_{0}^{c}}\|_{1}\neq 0$, the left-hand side of (\ref{eq:NoiselessCondition}) must be negative, which is a contradiction. We thus conclude that $\hat{\boldsymbol{\beta}}_{S_{0}^{c}}=\boldsymbol{0}$, and the KKT conditions (\ref{eq:NoiselessKKT1}) and (\ref{eq:NoiselessKKT2}) reduce to
\begin{align}\label{eq:NoiselessKKT3}
2 \mathbf{C}_{ww}(S_{0},S_{0}) \left(\hat{\boldsymbol{\beta}}_{S_{0}}-\boldsymbol{\beta}_{S_{0}}^{0}\right) + 2 \mathbf{C}_{wu}(S_{0},S_{0}) \boldsymbol{\beta}_{S_{0}}^{0} = -\lambda\hat{\boldsymbol{\tau}}_{S_{0}} \\ \label{eq:NoiselessKKT4}
2 \mathbf{C}_{ww}(S_{0}^{c},S_{0}) \left(\hat{\boldsymbol{\beta}}_{S_{0}}-\boldsymbol{\beta}_{S_{0}}^{0}\right)  + 2 \mathbf{C}_{wu}(S_{0}^{c},S_{0}) \boldsymbol{\beta}_{S_{0}}^{0} = -\lambda\hat{\boldsymbol{\tau}}_{S_{0}^{c}},
\end{align}
From (\ref{eq:NoiselessKKT3}) we get
\begin{align}\label{eq:NoiselessCondition2}
\left|\hat{\boldsymbol{\beta}}_{S_{0}} - \boldsymbol{\beta}_{S_{0}}^{0}\right| = \left|\frac{\lambda}{2} \mathbf{C}_{S_{0}}(S_{0},S_{0})^{-1}\hat{\boldsymbol{\tau}}_{S_{0}} + \mathbf{C}_{ww}(S_{0},S_{0})^{-1} \mathbf{C}_{wu}(S_{0},S_{0}) \boldsymbol{\beta}_{S_{0}}^{0} \right| \\ \nonumber
\leq \left(\frac{\lambda}{2} \underset{\|\boldsymbol{\tau}\|_{\infty}\leq 1}{\text{sup}} \left\| \mathbf{C}_{ww}(S_{0},S_{0})^{-1}\boldsymbol{\tau}_{S_{0}}  \right\|_{\infty} \right)\boldsymbol{1}+\left| \mathbf{C}_{ww}(S_{0},S_{0})^{-1} \mathbf{C}_{wu}(S_{0},S_{0})\boldsymbol{\beta}_{S_{0}}^{0} \right| .
\end{align}
Now, if $j \in S_{0}^{\text{det}}$ and $\hat{\beta}_{j}=0$, then
\begin{align*}
&\left|\hat{\beta}_{j} - \beta_{j}^{0}\right| = \left|\beta_{j}^{0}\right| > \frac{\lambda}{2} +\left(\underset{\|\boldsymbol{\tau}_{\infty}\|\leq 1}{\text{sup}} \left\| \mathbf{C}_{ww}(S_{0},S_{0})^{-1}\boldsymbol{\tau}_{S_{0}}  \right\|_{\infty}\right) +\left| v_{j} \right| ,
\end{align*}
where 
\begin{equation*}
\mathbf{v} = \left(v_{1},\dots,v_{p}\right)' = \mathbf{C}_{ww}(S_{0},S_{0})^{-1} \mathbf{C}_{wu}(S_{0},S_{0})\boldsymbol{\beta}_{S_{0}}^{0},
\end{equation*}
contradicting (\ref{eq:NoiselessCondition2}). Thus, $\hat{\beta}_{j}\neq 0$ for $j \in S_{0}^{\text{det}}$.

\subsubsection*{Part 2}
We start by assuming $\text{sign}(\hat{\boldsymbol{\beta}}) = \text{sign}(\boldsymbol{\beta}^{0})$. Thus, the KKT conditions are (\ref{eq:NoiselessKKT3}) and (\ref{eq:NoiselessKKT4}). From (\ref{eq:NoiselessKKT3}) we get
\begin{align*}
\hat{\boldsymbol{\beta}}_{S_{0}} - \boldsymbol{\beta}_{S_{0}}^{0} = -\frac{\lambda}{2}\mathbf{C}_{ww}(S_{0},S_{0})^{-1} \hat{\boldsymbol{\tau}}_{S_{0}} - \mathbf{C}_{ww}(S_{0},S_{0})^{-1} \mathbf{C}_{wu}(S_{0},S_{0})\boldsymbol{\beta}_{S_{0}}^{0}.
\end{align*}
Inserting this into (\ref{eq:NoiselessKKT4}) yields
\begin{align*}
&\mathbf{C}_{ww}(S_{0}^{c},S_{0}) \mathbf{C}_{ww}(S_{0},S_{0})^{-1} \hat{\boldsymbol{\tau}}_{S_{0}} + \\
&\frac{2}{\lambda}\left( \mathbf{C}_{ww}(S_{0}^{c},S_{0}) \mathbf{C}_{ww}(S_{0},S_{0})^{-1} \mathbf{C}_{wu}(S_{0},S_{0}) - \mathbf{C}_{wu}(S_{0}^{c},S_{0})\right)\boldsymbol{\beta}_{S_{0}}^{0} = \hat{\boldsymbol{\tau}}_{S_{0}^{c}},
\end{align*}
and the necessary condition stated in Proposition 3 follows by definition.

\par

\section{Proof of Theorem 2}
\setcounter{equation}{0}
Starting from the KKT conditions of Lemma 2, we will redo the steps of the proof of Theorem 1, but with the insertion of extra terms representing the correction for measurement error. The corrected lasso is not in general convex, and our analysis will thus concern \emph{any} critical point $\hat{\boldsymbol{\gamma}} = \hat{\boldsymbol{\beta}} - \boldsymbol{\beta}^{0}$ in the interior of the feasible set $\{\boldsymbol{\gamma} : \|{\boldsymbol{\gamma}} + \boldsymbol{\beta}^{0}\| < R\}$. 

If $\hat{\boldsymbol{\gamma}}$ exists, and satisfies the three conditions
\begin{align}\label{eq:KKTCorrected1}
&- \frac{\mathbf{W}_{S_{0}}'}{\sqrt{n}} \boldsymbol{\epsilon} + \sqrt{n} \left(\mathbf{C}_{ww}\left(S_{0},S_{0} \right)-\boldsymbol{\Sigma}_{uu}\left(S_{0},S_{0}\right)\right) \hat{\boldsymbol{\gamma}}_{S_{0}} +\\ \nonumber
&\qquad \sqrt{n} \left(C_{wu}\left(S_{0},S_{0} \right) -\boldsymbol{\Sigma}_{uu}\left(S_{0},S_{0}\right)\right)\boldsymbol{\beta}_{S_{0}}^{0}
 = - \frac{\lambda\sqrt{n}}{2 } \text{sign}\left(\boldsymbol{\beta}^{0}_{S_{0}}\right) \\ \label{eq:KKTCorrected2}
&\left|\hat{\boldsymbol{\gamma}}_{S_{0}} \right| < \left|\boldsymbol{\beta}_{S_{0}}^{0} \right| \\ \label{eq:KKTCorrected3}
&\bigg|- \frac{W_{S_{0}^{c}}'}{\sqrt{n}} \boldsymbol{\epsilon} + \sqrt{n} \left(\mathbf{C}_{ww}\left( S_{0}^{c},S_{0}\right)- \boldsymbol{\Sigma}_{uu}\left(S_{0}^{c},S_{0}\right)\right) \hat{\boldsymbol{\gamma}}_{S_{0}} + \\ \nonumber
&\qquad \sqrt{n} \left(\mathbf{C}_{wu}\left(S_{0}^{c},S_{0} \right)- \boldsymbol{\Sigma}_{uu}\left(S_{0}^{c},S_{0}\right)\right) \boldsymbol{\beta}_{S_{0}}^{0} \bigg| \leq \frac{\lambda \sqrt{n}}{2} \boldsymbol{1},
\end{align}
then $\text{sign}(\hat{\boldsymbol{\beta}}_{S_{0}}) = \text{sign}(\boldsymbol{\beta}^{0}_{S_{0}})$ and $\text{sign}(\hat{\boldsymbol{\beta}}_{S_{0}})  = \mathbf{0}$.

Event $A$ in Theorem 2 implies the existence of $|\hat{\boldsymbol{\gamma}}_{S_{0}}| < |\boldsymbol{\beta}_{S_{0}}^{0}|$ such that
\begin{align*}
\left|\mathbf{Z}_{6} - \mathbf{Z}_{7}\boldsymbol{\beta}_{S_{0}}\right| = \sqrt{n}\left( \left|\hat{\boldsymbol{\gamma}}_{S_{0}}\right| - \frac{\lambda}{2}\left|\left(\mathbf{C}_{ww}\left(S_{0},S_{0}\right) - \boldsymbol{\Sigma}_{uu}\left(S_{0},S_{0}\right)\right)^{-1} \text{sign}\left(\boldsymbol{\beta}_{S_{0}}^{0}\right)\right|\right).
\end{align*}
But then there must exist $|\hat{\boldsymbol{\gamma}}_{S_{0}}| < |\boldsymbol{\beta}_{S_{0}}^{0}|$ such that
\begin{align*}
\mathbf{Z}_{6} - \mathbf{Z}_{7}\boldsymbol{\beta}_{S_{0}} = \sqrt{n}\left( \hat{\boldsymbol{\gamma}}_{S_{0}}- \frac{\lambda}{2}\left(\mathbf{C}_{ww}\left(S_{0},S_{0}\right) - \boldsymbol{\Sigma}_{uu}\left(S_{0},S_{0}\right)\right)^{-1} \text{sign}\left(\boldsymbol{\beta}_{S_{0}}^{0}\right)\right).
\end{align*}
Multiplying through by $\mathbf{C}_{ww}(S_{0},S_{0}) - \boldsymbol{\Sigma}_{uu}(S_{0},S_{0})$ and reorganizing terms, we get (\ref{eq:KKTCorrected1}). Thus, $A$ ensures that (\ref{eq:KKTCorrected1}) and (\ref{eq:KKTCorrected2}) are satisfied. Next, adding and subtracting
\begin{align*}
\sqrt{n}\left(\mathbf{C}_{ww}\left(S_{0}^{c},S_{0}\right) - \boldsymbol{\Sigma}_{uu} \left(S_{0}^{c},S_{0}\right)\right)\hat{\boldsymbol{\gamma}}_{S_{0}}
\end{align*} 
to the left-hand side of event $B$ and the using the triangle inequality, yields
\begin{align*}
&\left| -\frac{\mathbf{W}_{S_{0}^{c}}'}{\sqrt{n}} \boldsymbol{\epsilon} + \sqrt{n} \left(\left( \mathbf{C}_{wu}\left(S_{0}^{c},S_{0}\right) - \boldsymbol{\Sigma}_{uu}\left(S_{0}^{c},S_{0}\right) \right)\hat{\boldsymbol{\gamma}}_{S_{0}}+ \left( \mathbf{C}_{wu}\left(S_{0}^{c},S_{0} \right) -\boldsymbol{\Sigma}_{uu}\left(S_{0}^{c},S_{0}\right)\right) \boldsymbol{\beta}_{S_{0}}^{0}\right)\right|\\
& \bigg| -\left(\mathbf{C}_{ww}\left(S_{0}^{c},S_{0}\right) - \mathbf{\Sigma}_{uu}\left(S_{0}^{c},S_{0}\right)\right) \left( \mathbf{C}_{ww}\left(S_{0},S_{0}\right) - \mathbf{\Sigma}_{uu}\left(S_{0},S_{0}\right)\right)^{-1}\frac{\mathbf{W}_{S_{0}}'}{\sqrt{n}} + \\
& \sqrt{n} \left( \mathbf{C}_{ww}\left(S_{0}^{c},S_{0}\right) - \mathbf{\Sigma}_{uu}\left(S_{0}^{c},S_{0}\right)\right) \left( \mathbf{S}_{ww}\left(S_{0},S_{0}\right) - \mathbf{\Sigma}_{uu}\left(S_{0},S_{0}\right)\right)^{-1} \\
& \left( \mathbf{C}_{wu}\left(S_{0},S_{0}\right) - \mathbf{\Sigma}_{uu}\left(S_{0},S_{0}\right)\right) \boldsymbol{\beta}_{S_{0}}^{0} + \sqrt{n}\left( \mathbf{S}_{ww}\left(S_{0}^{c},S_{0}\right) - \mathbf{\Sigma}_{uu}\left(S_{0}^{c},S_{0}\right)\right) \hat{\boldsymbol{\gamma}}_{S_{0}} 
 \bigg| \\
 &\leq \frac{\lambda\sqrt{n}}{2} \left(1-\theta\right) \mathbf{1}.
\end{align*}
The second term on the left-hand side of this expression is the left-hand side of (\ref{eq:KKTCorrected1}) multiplied by 
\begin{align*}
\left(\mathbf{C}_{ww}\left(S_{0}^{c},S_{0}\right) - \mathbf{\Sigma}_{uu}\left(S_{0}^{c},S_{0}\right)\right) \left( \mathbf{C}_{ww}\left(S_{0},S_{0}\right) - \mathbf{\Sigma}_{uu}\left(S_{0},S_{0}\right)\right)^{-1}.
\end{align*}
It can thus be replaced by the right-hand side of (\ref{eq:KKTCorrected1}) multiplied by this factor. This yields
\begin{align*}
&\left| -\frac{\mathbf{W}_{S_{0}^{c}}'}{\sqrt{n}} \boldsymbol{\epsilon} + \sqrt{n} \left(\left( \mathbf{C}_{wu}\left(S_{0}^{c},S_{0}\right) - \boldsymbol{\Sigma}_{uu}\left(S_{0}^{c},S_{0}\right) \right)\hat{\boldsymbol{\gamma}}_{S_{0}}+ \left( \mathbf{C}_{wu}\left(S_{0}^{c},S_{0} \right) -\boldsymbol{\Sigma}_{uu}\left(S_{0}^{c},S_{0}\right)\right) \boldsymbol{\beta}_{S_{0}}^{0}\right)\right|\\
& \left| \frac{\lambda \sqrt{n}}{2}\left(\mathbf{C}_{ww}\left(S_{0}^{c},S_{0}\right) - \mathbf{\Sigma}_{uu}\left(S_{0}^{c},S_{0}\right)\right) \left( \mathbf{C}_{ww}\left(S_{0},S_{0}\right) - \mathbf{\Sigma}_{uu}\left(S_{0},S_{0}\right)\right)^{-1} \text{sign}\left(\boldsymbol{\beta}_{S_{0}}^{0}\right)
 \right| \\
 &\leq \frac{\lambda\sqrt{n}}{2} \left(1-\theta\right) \mathbf{1},
\end{align*}
which implies, due to the IC-CL,
\begin{align*}
&\bigg| -\frac{\mathbf{W}_{S_{0}^{c}}'}{\sqrt{n}} \boldsymbol{\epsilon} + \sqrt{n} \left( \mathbf{C}_{wu}\left(S_{0}^{c},S_{0}\right) - \boldsymbol{\Sigma}_{uu}\left(S_{0}^{c},S_{0}\right) \right)\hat{\boldsymbol{\gamma}}_{S_{0}} \\
& \qquad + \sqrt{n}\left( \mathbf{C}_{wu}\left(S_{0}^{c},S_{0} \right) -\boldsymbol{\Sigma}_{uu}\left(S_{0}^{c},S_{0}\right)\right) \boldsymbol{\beta}_{S_{0}}^{0}\bigg|\leq \frac{\lambda\sqrt{n}}{2}  \mathbf{1}.
\end{align*}
This is indeed (\ref{eq:KKTCorrected3}). Altogether, $A$ implies (\ref{eq:KKTCorrected1}) and (\ref{eq:KKTCorrected2}), while $B|A$ implies (\ref{eq:KKTCorrected3}).

For the asymptotic result, define the vectors
\begin{align*}
&\mathbf{z} = \left(\mathbf{C}_{ww}(S_{0},S_{0}) - \boldsymbol{\Sigma}_{uu}\left(S_{0},S_{0}\right)\right)^{-1}  \frac{\mathbf{W}_{S_{0}}'}{\sqrt{n}}\boldsymbol{\epsilon} \\
&\mathbf{a} = \left| \boldsymbol{\beta}_{S_{0}}^{0}\right| - \left|\left(\mathbf{C}_{ww}(S_{0},S_{0}) - \boldsymbol{\Sigma}_{uu}(S_{0},S_{0}) \right)^{-1} \left(\mathbf{C}_{wu}(S_{0},S_{0}) - \boldsymbol{\Sigma}_{uu}(S_{0},S_{0})\right) \boldsymbol{\beta}_{S_{0}}^{0} \right| \\
&\mathbf{b} =\left( \mathbf{C}_{ww}(S_{0},S_{0}) - \boldsymbol{\Sigma}_{uu}(S_{0},S_{0}) \right)^{-1} \text{sign}\left(\boldsymbol{\beta}_{S_{0}}^{0} \right) \\
&\boldsymbol{\zeta} = \left( \left(\mathbf{C}_{ww}(S_{0}^{c},S_{0}) - \boldsymbol{\Sigma}_{uu}(S_{0}^{c},S_{0})\right) \left(\mathbf{C}_{ww}(S_{0},S_{0}) - \boldsymbol{\Sigma}_{uu}(S_{0},S_{0})\right)^{-1} \frac{\mathbf{W}_{S_{0}}'}{\sqrt{n}} - \frac{\mathbf{W}_{S_{0}^{c}}'}{\sqrt{n}} \right)\boldsymbol{\epsilon} \\
&\mathbf{f} = \bigg( \left(\mathbf{C}_{ww}(S_{0}^{c},S_{0}) - \boldsymbol{\Sigma}_{uu}(S_{0}^{c},S_{0})\right)\left(\mathbf{C}_{ww}(S_{0},S_{0}) - \boldsymbol{\Sigma}_{uu}(S_{0},S_{0})\right)^{-1} \\
&\qquad \left( \mathbf{C}_{wu}(S_{0},S_{0}) - \boldsymbol{\Sigma}_{uu}(S_{0},S_{0})\right)- \left(\mathbf{C}_{wu}(S_{0}^{c},S_{0})- \boldsymbol{\Sigma}_{uu}(S_{0}^{c},S_{0})\right)\bigg)\boldsymbol{\beta}_{S_{0}}^{0}.
\end{align*}
We have
\begin{align*}
&1- P(A\cap B) \leq P(A^{c}) +P(B^{c}) 
\leq \\
&\qquad \sum_{j=1}^{s_{0}} P\left(\left|z_{j} \right| \geq \sqrt{n}\left(a_{j} - \frac{\lambda}{2} b_{j} \right)\right) + \sum_{j=1}^{p-s_{0}}P\left(\left| \zeta_{j} - \sqrt{n}f_{j}\right| \geq \frac{\lambda \sqrt{n}}{2} \left(1-\theta\right) \right).
\end{align*}
It is clear that 
\begin{align*}
\mathbf{z} \overset{d}{\to} \mathcal{N}\left(\mathbf{0}, \sigma^{2} \boldsymbol{\Sigma}_{xx}\left(S_{0},S_{0}\right)^{-1} \boldsymbol{\Sigma}_{ww}\left(S_{0},S_{0}\right)\boldsymbol{\Sigma}_{xx}\left(S_{0},S_{0}\right)^{-1} \right), \text{ as } n\to \infty.
\end{align*}
Hence, there exists a finite constant $k$ such that $E(z_{j})^{2} < k^{2}$ for $j=1,\dots,s_{0}$. Next, we have by assumption
\begin{align*}
\mathbf{a} \to \left|\boldsymbol{\beta}_{S_{0}}^{0}\right|, \text{ as } n \to \infty.
\end{align*}
Now using the assumption $\lambda = o(1)$, we get
\begin{align*}
P\left(A^{c}\right) &\leq \sum_{j=1}^{s_{0}} \left(1 - P\left(\frac{\left|z_{j}\right|}{k} < \frac{\sqrt{n}}{2k}a_{j}\left(1+o(1)\right) \right) \right)\\
&\leq \left(1 + o(1) \right)\sum_{j=1}^{s_{0}} \left(1 - \Phi\left(\frac{\sqrt{n}}{2s}a_{j} \left(1+o(1)\right) \right) \right) \\
&= o\left(\exp(-n^{c})\right),
\end{align*}
where we used the bound (\ref{eq:GaussianTail}). Next, we note that
\begin{align*}
\zeta \overset{d}{\to} \mathcal{N}\left(\mathbf{0}, \sigma^{2}\left(\boldsymbol{\Sigma}_{xx}(S_{0}^{c},S_{0}^{c}) - \boldsymbol{\Sigma}_{xx}(S_{0}^{c},S_{0}) \boldsymbol{\Sigma}_{xx}(S_{0},S_{0})^{-1} \boldsymbol{\Sigma}_{xx}(S_{0},S_{0}^{c})\right) \right), \text{ as } n\to \infty.
\end{align*}
Next, we consider $\mathbf{f}$, and note that the limiting distribution of
\begin{align*}
\sqrt{n} \left( \mathbf{C}_{wu} - \boldsymbol{\Sigma}_{uu}\right), \text{ as } n\to \infty,
\end{align*}
is normal with mean $\mathbf{0}$ and finite variances (Anderson (2003, Th. 3.4.4)). In addition,
\begin{align*}
\mathbf{C}_{ww} - \boldsymbol{\Sigma}_{uu} \to \boldsymbol{\Sigma}_{xx}, \text{ as } n \to \infty.
\end{align*}
Thus, applying Slutsky's theorem to the product of the matrices, the limiting distribution of
\begin{align*}
&\sqrt{n} \bigg( \left( \mathbf{C}_{ww}(S_{0}^{c},S_{0}) - \boldsymbol{\Sigma}_{uu}(S_{0}^{c},S_{0})\right) \left( \mathbf{C}_{ww}(S_{0},S_{0}) - \boldsymbol{\Sigma}_{uu}(S_{0},S_{0}) \right)^{-1} \\
&\qquad \left( \mathbf{C}_{wu}(S_{0},S_{0}) - \boldsymbol{\Sigma}_{uu}(S_{0},S_{0})\right) - \left( \mathbf{C}_{wu}(S_{0}^{c},S_{0}) - \boldsymbol{\Sigma}_{uu}(S_{0}^{c},S_{0})\right) \bigg), \text{ as } n\to \infty,
\end{align*}
is normal with mean $\mathbf{0}$ and finite variances. Now,
\begin{align*}
&\sqrt{n} \mathbf{f}  =\\
&\sqrt{n} \bigg( \left( \mathbf{C}_{ww}(S_{0}^{c},S_{0}) - \boldsymbol{\Sigma}_{uu}(S_{0}^{c},S_{0})\right) \left( \mathbf{C}_{ww}(S_{0},S_{0}) - \boldsymbol{\Sigma}_{uu}(S_{0},S_{0}) \right)^{-1} \\
&\qquad \left( \mathbf{C}_{wu}(S_{0},S_{0}) - \boldsymbol{\Sigma}_{uu}(S_{0},S_{0})\right) - \left( \mathbf{C}_{wu}(S_{0}^{c},S_{0}) - \boldsymbol{\Sigma}_{uu}(S_{0}^{c},S_{0})\right) \bigg) \boldsymbol{\beta}_{S_{0}}^{0}
\end{align*}
is a vector in $\mathbb{R}^{p-s_{0}}$ whose elements are linear combinations of variables whose limiting distribution as $n \to \infty$ is normal with mean zero and finite variances. Accordingly, the limiting distribution of $\sqrt{n}\mathbf{f}$ as $n \to \infty$ is normal with mean zero and finite variances.

So again there exists a finite constant $k$ such that $E(\zeta_{j} - \sqrt{n}f_{j})^{2} < k^{2}$ for $j=1,\dots,(p-s_{0})$. Thus, when $\lambda n^{(1-c)/2} \to \infty$ for $c \in [0,1)$, we have
\begin{align*}
P(B^{c}) &\leq \sum_{j=1}^{p-s_{0}} \left(1 - P\left(\frac{\left|\zeta_{j} - \sqrt{n}f_{j} \right|}{k} < \frac{1}{k} \frac{\lambda \sqrt{n}}{2}\left( 1 - \theta\right) \right) \right) \\
&\leq \left( 1 + o(1)\right) \sum_{j=1}^{p-s_{0}} \left(1 - \Phi\left(\frac{1}{k} \frac{\lambda \sqrt{n}}{2} \left( 1 - \theta\right) \right) \right)\\
& = o\left( \exp(-n^{c})\right).
\end{align*}
It follows that $P(A \cap B) = 1 - o(\exp(-n^{c}))$.
\\
\par

\noindent{\large\bf Additional References}
\begin{description}
\item
Anderson, T. W. (2003). {\it An introduction to multivariate statistical analysis, third edition}. {John Wiley and Sons, Hoboken.}
\end{description}
\end{document}